\documentclass[preprintnumbers]{revtex4}
\usepackage{amsfonts}
\usepackage{amsmath}
\usepackage{hyperref}
\usepackage{amssymb}
\usepackage{subfigure}
\usepackage[english]{babel}
\usepackage{graphicx}
\usepackage{epsfig}
\usepackage{bm}
\usepackage{longtable}
\usepackage{verbatim}
\usepackage{longtable}
\usepackage[utf8]{inputenc}

\newcommand{\mathsym}[1]{{}}

\input{tcilatex}
\begin{document}

\title{Radiative Seesaw-type Mechanism of Fermion Masses and Non-trivial
Quark Mixing}
\author{Carolina Arbel\'aez}
\email{carolina.arbelaez@usm.cl}
\author{A. E. C\'arcamo Hern\'andez}
\email{antonio.carcamo@usm.cl}
\author{Sergey Kovalenko}
\email{sergey.kovalenko@usm.cl}
\author{Ivan Schmidt}
\email{ivan.schmidt@usm.cl}
\affiliation{{\small Centro Cient\'ifico-Tecnol\'ogico de Valpara\'{\i}so-CCTVal,
Universidad T\'ecnica Federico Santa Mar\'ia,}\\
Casilla 110-V, Valpara\'{\i}so, Chile}
\date{\today }

\begin{abstract}
We propose a predictive inert two Higgs doublet model, where the standard
Model (SM) symmetry is extended by $S_{3}\otimes Z_{2}\otimes Z_{12}$ and
the field content is enlarged by extra scalar fields, charged exotic
fermions and two heavy right-handed Majorana neutrinos. The charged exotic
fermions generate a non-trivial quark mixing and provide one-loop-level
masses for the first- and second-generation charged fermions. The masses of
the light active neutrinos are generated from a one-loop-level radiative
seesaw mechanism. Our model successfully explains the observed SM fermion
mass and mixing pattern. 
\end{abstract}

\maketitle



\section{Introduction}

Despite its great success, the standard Model (SM) does not address several
fundamental issues such as, for example, the number of fermion families and
the observed pattern of fermion masses and mixing. As is known, in the quark
sector the mixing is small while in the lepton sector two of the mixing
angles are large. 
The three neutrino flavors mix with each other and at least two of the
neutrinos have non-vanishing masses, which according to neutrino oscillation
experimental data must be smaller than the SM charged fermion masses by many
orders of magnitude. This so called ``flavor puzzle'' 
motivates extensions of the SM with larger scalar and/or fermion sector and
with extended gauge groups, supplemented with discrete flavor symmetries, so
that the resulting 
fermion mass matrix textures would explain the observed pattern of fermion
masses and mixing. The implementation of these discrete flavor symmetries in
several extensions of the SM is expected to 
provide 
an elegant solution of the ``flavor puzzle'' (for recent reviews on discrete
flavor groups see, for instance, Refs.~\cite%
{Ishimori:2010au,Altarelli:2010gt,King:2013eh,King:2014nza}). 
In fact, considerable 
attention has already been paid in the literature to the $S_{3}$ \cite%
{Pakvasa:1977in,Cardenas:2012bg,Dias:2012bh,Dev:2012ns,Meloni:2012ci,Canales:2013cga,Ma:2013zca,Kajiyama:2013sza,Hernandez:2013hea,Hernandez:2014lpa,Hernandez:2014vta,Vien:2014vka,Ma:2014qra,Das:2015sca,Hernandez:2015dga,Hernandez:2015zeh,Hernandez:2016rbi,Hernandez:2015hrt,CarcamoHernandez:2016pdu,Xing:2010iu}
$A_{4}$ \cite%
{Ma:2001dn,Babu:2002dz,Altarelli:2005yp,Altarelli:2005yx,deMedeirosVarzielas:2005qg,He:2006dk,Varzielas:2012ai,Cooper:2012wf,Ishimori:2012fg,Ahn:2013mva,Memenga:2013vc,Bhattacharya:2013mpa,Ferreira:2013oga,Felipe:2013vwa,Hernandez:2013dta,King:2013hj,Morisi:2013qna,Morisi:2013eca,Felipe:2013ie,King:2014iia,Karmakar:2014dva,Campos:2014lla,Hernandez:2015tna,Karmakar:2015jza,Pramanick:2015qga,Bonilla:2016sgx,Belyaev:2016oxy,CarcamoHernandez:2017cwi,CarcamoHernandez:2017kra}%
, $S_{4}$ \cite%
{Yang:2011fh,BhupalDev:2012nm,Mohapatra:2012tb,Varzielas:2012pa,Ding:2013hpa,Ishimori:2010fs,Ding:2013eca,Hagedorn:2011un,Campos:2014zaa}%
, $T_{7}$ \cite%
{Luhn:2007sy,Hagedorn:2008bc,Cao:2010mp,Luhn:2012bc,Kajiyama:2013lja,Bonilla:2014xla,Hernandez:2015cra,Arbelaez:2015toa,Hernandez:2015yxx}
and $\Delta (27)$ \cite%
{deMedeirosVarzielas:2006fc,Ma:2006ip,Varzielas:2012nn,Bhattacharyya:2012pi,Ferreira:2012ri,Ma:2013xqa,Nishi:2013jqa,Varzielas:2013sla,Aranda:2013gga,Varzielas:2013eta,Abbas:2014ewa,Varzielas:2015aua,Bjorkeroth:2015uou,Chen:2015jta,Abbas:2015zna,Vien:2016tmh,Hernandez:2016eod,Chulia:2016giq,CarcamoHernandez:2017owh}
flavor groups. Several models with discrete flavor symmetries, which bring
about radiative seesaw mechanisms of fermion mass generation, have also been discussed in the literature \cite%
{Ma:1989ys,Ma:1989cn,Dong:2006gx,Hernandez:2013mcf,Adhikari:2015woo,Nomura:2016emz,Kownacki:2016hpm,Nomura:2016ezz}%
. A typical flaw of these models is the large number of free parameters and
consequent limited predictive power, since they lack a mechanism underlying
the observed hierarchy of quark mixing angles and masses. The experimental
data suggest an empirical relation of the Cabbibo mixing angle to the first-
and second-generation down-type quark masses $\sin\theta_C\sim\sqrt{\frac{m_d%
}{m_s}}$ \cite{Gatto:1968ss,Cabibbo:1968vn,Oakes:1969vm}, which seems to
imply a radiative seesaw mechanism of fermion mass generation, where the
Cabbibo mixing arises from the down-type quark sector, whereas the up quark
sector contributes to the remaining mixing angles. Inspired with this
observation we propose an extension of the inert two Higgs doublet model
(2HDM) realizing this idea. 
In our model a mismatch between the down- and up-type quark mass matrix
textures 
is introduced, by distinguishing the two Higgs doublets with respect to the $%
Z_{2}$ flavor symmetry, 
preserved at all scales. One of these doublets--the SM Higgs--is $Z_{2}$%
-even, while another one is $Z_{2}$-odd. In order to maintain the $Z_{2}$%
-symmetry intact, we assume the latter to be VEV-less. The described setup
is similar to the well-known inert 2HDM \cite{Deshpande:1977rw}, but with
the difference that here some of the SM fermions are $Z_{2}$-odd. %
%
Furthermore, compared to Ref.~\cite{Deshpande:1977rw}, the discrete symmetry
in our model is extended from $Z_{2}$ to $S_{3}\otimes Z_{2}\otimes Z_{12}$,
and the field content is enlarged by extra scalars, heavy charged exotic
fermions and two heavy right-handed Majorana neutrinos. Note that only the $%
Z_{2}$ is preserved at low energies, whereas the $S_{3}\otimes Z_{12}$
symmetry is spontaneously broken. The latter allows us to explain the
observed SM fermion mass and mixing pattern. 
In our modified inert 2HDM, the SM fermion mass and mixing pattern is due to
a combination of tree and 1-loop-level effects. At tree level only the third
generation charged fermions acquire masses and there is no quark mixing,
while the first and second generation charged fermion masses and the quark
mixing arise from one-loop-level radiative seesaw-type mechanism, triggered
by virtual $Z_{2}$-charged scalar fields and electrically charged exotic
fermions running inside %
the loops. Light active neutrino masses are generated from a one loop level
radiative seesaw mechanism. Due to the preserved $Z_{2}$ symmetry, our model
features natural dark matter candidates. 
The following comparison of our model with the similar models in the literature could be in order.
Despite the similar quality of the data description, our model is more
predictive than the model of Ref.~\cite{Branco:2010tx}, since the latter,
focused only on the quark sector, has a total of 12 free parameters, whereas
the quark sector of our model is described by 9 free effective parameters,
which are adjusted to reproduce the 10 physical observables of the quark
sector. The models of Refs.~\cite%
{Ishimori:2014jwa,Ishimori:2014nxa,Hernandez:2015zeh}, Refs.~\cite%
{CarcamoHernandez:2012xy,Campos:2014zaa,Hernandez:2015dga,Hernandez:2013hea,Hernandez:2014vta}%
, Refs.~\cite{Bhattacharyya:2012pi,Vien:2014ica} and Ref.~\cite%
{Branco:2011wz} possess in the quark sector 9, 10, 12, and 13 free
parameters, respectively. 
For the lepton sector, our model is less
predictive than in the quark sector, however, under some reasonable
assumptions, the number of effective parameters in this sector can be
reduced. As shown in detail in section III, our model can successfully
accommodate the eight physical observables in the lepton sector, with only five effective free parameters for the case of normal neutrino mass hierarchy.
With respect to the inverted neutrino mass hierarchy, more parameters are
needed to successfully reproduce the masses and leptonic mixing angles.

The content of this paper goes as follows. In Sect. \ref{themodel} we
introduce the model setup. Section \ref{fermionmassesandmixings} deals with
the derivation of fermion masses and mixings and provides our corresponding
results. Our conclusions are stated in Sect. \ref{conclusions}. Appendix %
\ref{S3} gives a brief description of the $S_{3}$ group. Appendix \ref{ep}
shows the analytical expressions for the dimensionless parameters of the SM
fermion mass matrices generated at one loop level. 

\section{The Model}

\label{themodel} 

We consider a modified inert 2HDM, with the Standard Model gauge symmetry
supplemented with the $S_{3}\otimes Z_{2}\otimes Z_{12}$ discrete group, and
a scalar sector composed of 2 scalar SM doublets, i.e., $\phi _{1}$ and $%
\phi _{2}$ plus 4 scalar SM singlets. Out of these four SM scalar singlets,
two scalar fields are grouped in a $S_{3}$ doublet, whereas the remaining
ones are assigned to be one $S_{3}$ trivial singlet and one $S_{3}$
non-trivial singlet. The $Z_{2}$ symmetry is assumed to be preserved whereas
the $S_{3}\otimes Z_{12}$ discrete group is broken at certain scale $\Lambda
_{int}$. The fermion sector of the SM is extended to include four SM gauge
singlet charged leptons $E_{1L}$, $E_{2L}$, $E_{1R}$ and $E_{2R}$, two right
handed neutrinos $N_{1R}$, $N_{2R}$ and four $SU\left( 2\right) _{L}$
singlet heavy quarks $T_{nL}$, $T_{nR}$, $B_{nL}$, $B_{nR}$ ($n=1,2$). The
reasons for these specific choices will be explained below. It is assumed
that the heavy exotic $T_{n}$ and $B_{n}$ quarks 
have electric charges equal to $\frac{2}{3}$ and $-\frac{1}{3}$,
respectively. The $S_{3}\otimes Z_{2}\otimes Z_{12}$ assignments of the
scalar fields are: 
\begin{eqnarray}
\phi _{1} &\sim &\left( \mathbf{1},1,1\right) ,\hspace{1.5cm}\phi _{2}\sim
\left( \mathbf{1,}-1,1\right) ,  \notag \\
\xi &\sim &\left( \mathbf{2},1,1\right) ,\hspace{1.5cm}\eta \sim \left( 
\mathbf{1},-1,1\right),\hspace{1.5cm}\tau\sim\left( \mathbf{1}^{\prime
},1,e^{-\frac{i\pi }{6}}\right) .  \label{scalarfields}
\end{eqnarray}%
The requirement of unbroken $Z_{2}$ symmetry implies that the $Z_{2}$
charged scalar fields $\phi _{2}$ and $\eta $ do not acquire vacuum
expectation values. The remaining scalar fields, i.e., $\phi _{1}$, $\xi $
and $\tau $, which are neutral under the $Z_{2}$ symmetry, have non
vanishing vacuum expectation values. Let us give a motivation for the above
presented extension (\ref{scalarfields}) of the scalar sector of the
original inert 2HDM \cite{Deshpande:1977rw}. 
One $Z_{2}$ odd SM singlet scalar $\eta $ (apart from the $Z_{2}$ odd scalar
doublet $\phi _{2}$) is needed to implement the radiative seesaw mechanism
of fermion mass generation. It yields a non-trivial quark mixing, provides
masses for the first- and second-generation charged fermions and contributes
to the light active neutrino masses. Furthermore, one $Z_{12}$-charged SM
singlet scalar ($\tau $) is needed to generate the SM charged fermion mass
and quark mixing pattern compatible with the observations. 
We also need an $S_{3}$-doublet scalar ($\xi $) to non-trivially couple it to
the first- and second-generation right-handed SM down-type quarks, which we
unify into an $S_{3}$-doublet. %

We decompose the Higgs doublets $\phi _{1}$ and $\phi _{2}$ 
in the standard way as 
\begin{eqnarray}
\phi _{l} &=&\left( 
\begin{array}{c}
\varphi _{l}^{+} \\ 
\frac{1}{\sqrt{2}}\left( v_{l}+\func{Re}\phi _{l}^{0}+i\func{Im}\phi
_{l}^{0}\right)%
\end{array}%
\right) ,  \notag \\
v_{l} &=&v\delta _{l1},\hspace{0.7cm}l=1,2.  \label{doublets}
\end{eqnarray}
Here $v$ is a Higgs VEV breaking the electroweak symmetry.

The $S_{3}$-symmetry is broken by the VEV of the $S_{3}$ scalar doublet $\xi$. 
We choose the VEV alignment
\begin{equation}
\left\langle \xi \right\rangle =v_{\xi }\left( 1,0\right), \label{VEV}
\end{equation}
compatible with the scalar potential minimization condition as demonstrated in 
Ref.~\cite{Hernandez:2015dga}. In the literature there have been constructed 
several $S_{3}$ flavor models with the similar VEV alignment of an $S_{3}$ scalar doublet (see for instance Refs.~\cite{Hernandez:2013hea,Hernandez:2014vta,Hernandez:2015dga,Hernandez:2015zeh}).

The $S_{3}\otimes Z_{2}\otimes Z_{12}$ assignments of the fermions of the
model are as follows 
\begin{eqnarray}
q_{1L} &\sim &\left( \mathbf{1}^{\prime }\mathbf{,}1,1\right) ,\hspace{1cm}%
q_{2L}\sim \left( \mathbf{1},1,1\right) ,\hspace{1cm}q_{3L}\sim \left( 
\mathbf{1},1,1\right) ,\hspace{1cm}  \notag \\
u_{3R} &\sim &\left( \mathbf{1},1,1\right) ,\hspace{1cm}u_{1R}\sim \left( 
\mathbf{1}^{\prime },-1,-1\right) ,\hspace{1cm}u_{2R}\sim \left( \mathbf{1}%
,-1,e^{\frac{i\pi }{3}}\right)  \notag \\
D_{R} &=&\left( d_{1R},d_{2R}\right) \sim \left( \mathbf{2},-1,e^{\frac{i\pi 
}{3}}\right) ,\hspace{1cm}d_{3R}\sim \left( \mathbf{1}^{\prime },1,i\right) ,
\notag \\
l_{1L} &\sim &\left( \mathbf{1},1,1\right) ,\hspace{1cm}l_{2L}\sim \left( 
\mathbf{1},1,1\right) ,\hspace{1cm}l_{3L}\sim \left( \mathbf{1},1,1\right) ,
\notag \\
l_{1R} &\sim &\left( \mathbf{1},-1,-1\right) ,\hspace{1cm}l_{2R}\sim \left( 
\mathbf{1}^{\prime },-1,i\right) ,\hspace{1cm}l_{3R}\sim \left( \mathbf{1}%
^{\prime },1,i\right) ,\hspace{1cm}  \notag \\
T_{1L} &\sim &\left( \mathbf{1},-1,1\right) ,\hspace{1cm}T_{1R}\sim \left( 
\mathbf{1}^{\prime },-1,1\right) ,\hspace{1cm}T_{2L}\sim \left( \mathbf{1}%
,-1,1\right) ,\hspace{1cm}T_{2R}\sim \left( \mathbf{1},-1,1\right) ,  \notag
\\
B_{1L} &\sim &\left( \mathbf{1},1,1\right) ,\hspace{1cm}B_{1R}\sim \left( 
\mathbf{1}^{\prime }\mathbf{,}-1,1\right) ,\hspace{1cm}B_{2L}\sim \left( 
\mathbf{1}^{\prime },1,1\right) ,\hspace{1cm}B_{2R}\sim \left( \mathbf{1,}%
-1,1\right) ,  \notag \\
E_{1L} &\sim &\left( \mathbf{1},-1,1\right) ,\hspace{1cm}E_{1R}\sim \left( 
\mathbf{1,-}1,1\right) ,\hspace{1cm}E_{2L}\sim \left( \mathbf{1},-1,1\right)
,\hspace{1cm}E_{2R}\sim \left( \mathbf{1,-}1,1\right) ,  \notag \\
N_{1R} &\sim &\left( \mathbf{1},-1,1\right) ,\hspace{1cm}N_{2R}\sim \left( 
\mathbf{1},-1,1\right) .  \label{Fermionassignments}
\end{eqnarray}%
%
%
%
%
%
%
%
%
%
%
%
%
%
%
%
%
%
%
%
%
%
%
%
%
%
%
%
%
Now it is timely to comment on the reasons for the introduction of the
extended symmetry $S_{3}\otimes Z_{2}\otimes Z_{12}$. The $S_{3}$ symmetry
reduces the number of parameters in the Yukawa sector of the model,
improving its predictivity. 
The $Z_{12}$ symmetry shapes the hierarchical structure of the fermion mass
matrices that gives rise to the observed charged fermion mass and quark
mixing pattern. The preserved $Z_{2}$ symmetry selects the allowed entries
in the quark mass matrices, so that the down-type quark sector contributes
to the Cabbibo mixing, whereas the up-type quark sector contributes to the
remaining mixing angles. 
The $Z_{2}$ symmetry allows also the implementation of a radiative seesaw
type-mechanism (induced by the $Z_{2}$ charged scalar fields and the exotic
charged fermions running in the internal lines of the loop), which gives
rise to a non-trivial quark mixing and generates the masses for the first- and
second-generation charged fermions. Besides that, due to the unbroken $Z_{2}$
symmetry, light active neutrino masses receive contributions from a one-loop
radiative seesaw mechanism (induced by the $\eta $ and $\phi_{2}$ scalar
fields and the heavy Majorana neutrinos $N_{1R}$ and $N_{2R}$ running in the
internal lines of the loops). Let us note that the masses of the light
active neutrinos are generated from a one-loop-level radiative seesaw
mechanism. It is worth mentioning that in order to be compatible with the
neutrino oscillation data, we need at least two light massive active
neutrinos. Having only one right-handed Majorana neutrino, will lead to two
massless active neutrinos, which is in clear contradiction with the
experimental data on neutrino oscillation experiments. That is why we
introduced in the model two massive right-handed neutrinos $N_{1R}$, $N_{2R}$, which is the minimal number necessary for this purpose, both for normal
and inverted neutrino mass hierarchy. For similar reasons we introduced in
our model four $SU\left( 2\right) _{L}$ singlet heavy quarks $T_{nL}$, $T_{nR}$, $B_{nL}$, $B_{nR}$ ($n=1,2$), necessary to avoid the appearance of
massless charged SM fermions.

With the above particle content, the following quark and lepton Yukawa
terms, invariant under the symmetries of the model, arise: 
\begin{eqnarray}
\tciLaplace _{Y}^{q} &=&m_{T_{1}}\overline{T}_{1L}T_{1R}+m_{T_{2}}\overline{T%
}_{2L}T_{2R}+y_{3}^{\left( u\right) }\overline{q}_{3L}\widetilde{\phi }%
_{1}u_{3R}  \notag \\
&&+y_{1}^{\left( u\right) }\overline{q}_{1L}\widetilde{\phi }%
_{2}T_{1R}+x_{1}^{\left( u\right) }\overline{T}_{1L}\eta u_{3R}  \notag \\
&&+y_{2}^{\left( u\right) }\overline{q}_{2L}\widetilde{\phi }%
_{2}T_{2R}+x_{2}^{\left( u\right) }\overline{T}_{2L}\eta u_{3R}  \notag \\
&&+x_{3}^{\left( u\right) }\overline{T}_{1L}\eta u_{1R}\frac{\tau ^{6}}{%
\Lambda ^{6}}+x_{4}^{\left( u\right) }\overline{T}_{2L}\eta u_{2R}\frac{\tau
^{2}}{\Lambda ^{2}}  \notag \\
&&+m_{B_{1}}\overline{B}_{1L}B_{1R}+m_{B_{2}}\overline{B}_{2L}B_{2R}+y_{3}^{%
\left( d\right) }\overline{q}_{3L}\phi _{1}d_{3R}\frac{\tau ^{3}}{\Lambda
^{3}}  \notag \\
&&+x_{2}^{\left( d\right) }\overline{B}_{1L}\eta D_{R}\frac{\xi \tau ^{2}}{%
\Lambda ^{3}}+x_{4}^{\left( d\right) }\overline{B}_{2L}\eta D_{R}\frac{\xi
\tau ^{2}}{\Lambda ^{3}}  \notag \\
&&+y_{1}^{\left( d\right) }\overline{q}_{1L}\phi _{2}B_{1R}+x_{1}^{\left(
d\right) }\overline{B}_{1L}\eta D_{R}\frac{\xi \xi \tau ^{2}}{\Lambda ^{4}} 
\notag \\
&&+y_{2}^{\left( d\right) }\overline{q}_{2L}\phi _{2}B_{2R}+x_{3}^{\left(
d\right) }\overline{B}_{2L}\eta D_{R}\frac{\xi \xi \tau ^{2}}{\Lambda ^{4}}%
+h.c  \label{Lyq}
\end{eqnarray}%
%
\begin{eqnarray}
\tciLaplace _{Y}^{l} &=&m_{E_{1}}\overline{E}_{1L}E_{1R}+m_{E_{2}}\overline{E%
}_{2L}E_{2R}+y_{3}^{\left( l\right) }\overline{l}_{3L}\phi _{1}l_{3R}\frac{%
\tau ^{3}}{\Lambda ^{3}}  \notag \\
&&+x_{1}^{\left( l\right) }\overline{l}_{1L}\phi _{2}E_{1R}+y_{1}^{\left(
l\right) }\overline{E}_{1L}\eta l_{1R}\frac{\tau ^{6}}{\Lambda ^{6}}  \notag
\\
&&+x_{2}^{\left( l\right) }\overline{l}_{2L}\phi _{2}E_{1R}+y_{2}^{\left(
l\right) }\overline{E}_{1L}\eta l_{2R}\frac{\tau ^{3}}{\Lambda ^{3}}  \notag
\\
&&+x_{3}^{\left( l\right) }\overline{l}_{1L}\phi _{2}E_{2R}+y_{3}^{\left(
l\right) }\overline{E}_{2L}\eta l_{1R}\frac{\tau ^{6}}{\Lambda ^{6}}  \notag
\\
&&+x_{4}^{\left( l\right) }\overline{l}_{2L}\phi _{2}E_{2R}+y_{4}^{\left(
l\right) }\overline{E}_{2L}\eta l_{2R}\frac{\tau ^{3}}{\Lambda ^{3}}  \notag
\\
&&+y_{11}^{\left( \nu \right) }\overline{l}_{1L}\widetilde{\phi }%
_{2}N_{1R}+y_{21}^{\left( \nu \right) }\overline{l}_{2L}\widetilde{\phi }%
_{2}N_{1R}  \notag \\
&&+y_{31}^{\left( \nu \right) }\overline{l}_{3L}\widetilde{\phi }%
_{2}N_{1R}+y_{12}^{\left( \nu \right) }\overline{l}_{1L}\widetilde{\phi }%
_{2}N_{2R}  \notag \\
&&+y_{22}^{\left( \nu \right) }\overline{l}_{2L}\widetilde{\phi }%
_{2}N_{2R}+y_{32}^{\left( \nu \right) }\overline{l}_{3L}\widetilde{\phi }%
_{2}N_{2R}  \notag \\
&&+m_{1N}\overline{N}_{1R}N_{1R}^{c}+m_{2N}\overline{N}_{2R}N_{2R}^{c}+h.c
\label{Lyl}
\end{eqnarray}%
Here we considered $S_{3}\otimes Z_{2}$ soft breaking mass terms for the
charged exotic fermions. In addition, we have neglected the mixings between
the $T_{1}$ ($B_{1}$) and $T_{2}$ ($B_{2}$) exotic quarks as well as the
mixings between the $E_{1}$ and $E_{2}$ charged exotic leptons, by
considering these charged exotic fermions as physical eigenstates. After the
spontaneous breaking of the electroweak and $S_{3}\otimes Z_{12}$ discrete
symmetries, these interactions generate at tree- and one-loop-levels the quark
and lepton mass matrices. 
The one loop Feynman diagrams contributions to the fermion mass matrices are
shown in Fig. 1. \newline
The hierarchy of charged fermion masses and quark mixing angles arises in
our model from the breaking of the $Z_{12}$ discrete group. In order to
relate the quark masses with the quark mixing parameters, the 
VEVs of the SM scalar singlets $\xi $ and $\tau $ are set as follows: 
\begin{equation}
v_{\xi }\sim \Lambda _{int}\sim v_{\tau }=\lambda \Lambda .
\end{equation}%

where $\lambda =0.225$ is one of the Wolfenstein parameters and $\Lambda $
corresponds to the model cutoff.\newline
\begin{figure*}[tbp]
\subfigure{\includegraphics[width=0.80\textwidth]{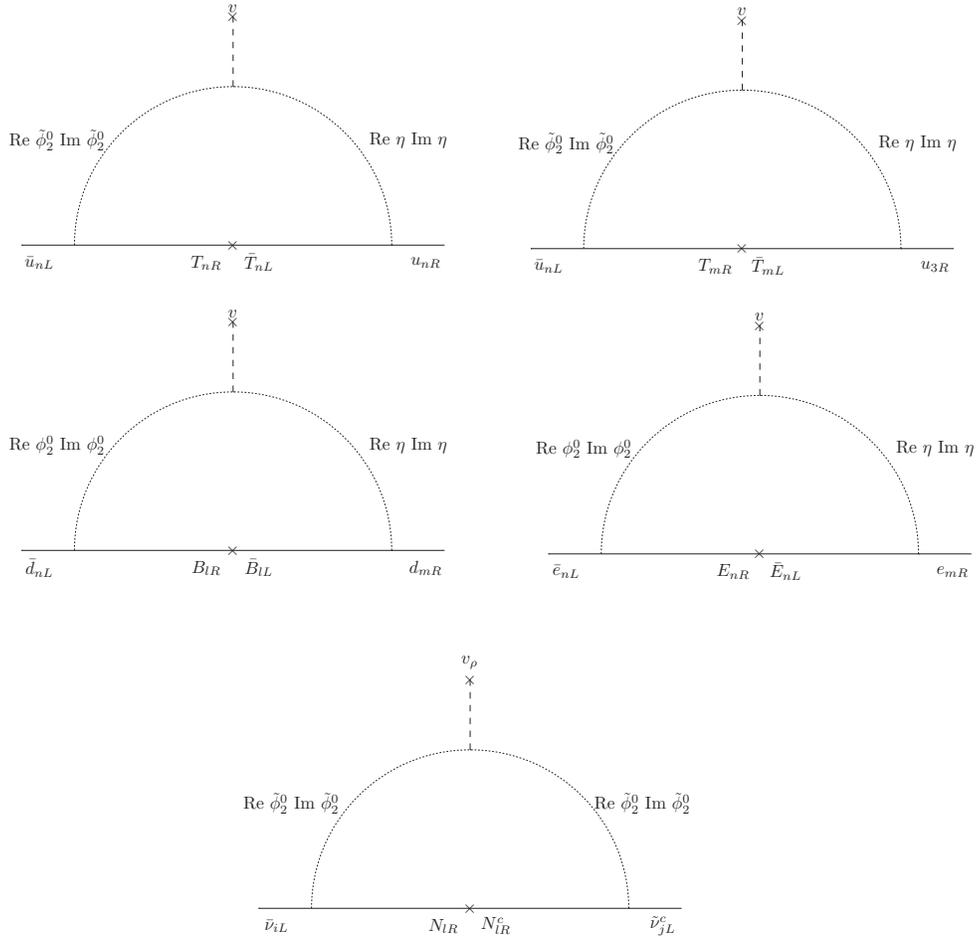}} 
\vspace{-5.0cm}
\caption{One loop Feynman diagrams contributions to the fermion mass
matrices. Here $n=1,2$, $m=1,2$ and $l=1,2$.}
\label{loopdiagrams}
\end{figure*}
From the quark Yukawa interactions it follows that the heavy exotic $T_{n}$ (%
$B_{n}$) quarks ($n=1,2$) will have a dominant decay mode into a SM up-
(down-) type quark and a heavy CP even or CP odd neutral Higgs boson, which
is identified as missing energy, due to the preserved $Z_{2}$ symmetry.
Furthermore, from the lepton Yukawa interactions it follows that the charged
exotic leptons $E_{n}$ ($n=1,2$) will decay dominantly into a SM charged
lepton and a heavy CP even or CP odd neutral Higgs boson. The exotic $T_{n}$
and $B_{n}$ quarks are produced in pairs at the LHC via gluon fusion and the
Drell-Yan mechanism, and the charged exotic leptons $E_{n}$ ($n=1,2$) are
also produced in pairs but only via the Drell-Yan mechanism. Thus, observing
an excess of events with respect to the SM background in the dijet and
opposite sign dileptons final states can be a signal in support of this
model at the LHC. On the other hand, it is worth mentioning that a heavy CP
even Higgs can be produced at the LHC in association with a CP odd Higgs
boson via Drell-Yan annihilation, with a cross section of about $0.1$ fb for
heavy CP even and CP odd scalar masses of $600$ GeV and LHC center of mass
energy of $\sqrt{s}=13$ TeV. To conclude this section, let us comment on the 
$h\rightarrow \gamma \gamma $ decay in our model, where $h$ is the $126$ GeV
Higgs boson. In the standard model, this decay is dominated by $W$ loop
diagrams, which can interfere destructively with the subdominant top quark
loop; whereas In the 2HDM, the $h\rightarrow \gamma \gamma $ decay receives
additional contributions from loops with charged scalars $H^{\pm }$,
proportional to $\sin 2\beta $ (for the case where the Higgs doublets have
different $Z_{N}$ charges), 
where $\tan \beta =v_{2}/v_{1}$ 
(cf. Ref.~\cite{Hernandez:2015dga}). In our model 
$\tan \beta =0$ since $v_{2}=0$, and thus the charged Higgs boson
contribution to the $126$~GeV Higgs diphoton decay is absent. From the
explicit form of the $h\rightarrow \gamma \gamma $ decay rate given in \cite%
{Hernandez:2015dga} for the 2HDM, it is easy to see that the Higgs diphoton
decay rate in our model coincides with the SM expectation, since the light
and heavy CP even neutral Higgs bosons do not mix, as a result of the
preserved $Z_{2}$ symmetry.

\section{Quark masses and mixings}
%
\label{fermionmassesandmixings} 
From the quark sector Yukawa terms (\ref{Lyq}) we find 
the quark mass matrices
\begin{eqnarray}
M_{U} &=&\left( 
\begin{array}{ccc}
\varepsilon _{11}^{\left( u\right) }\lambda ^{6} & 0 & \varepsilon
_{13}^{\left( u\right) } \\ 
0 & \varepsilon _{22}^{\left( u\right) }\lambda ^{2} & \varepsilon
_{23}^{\left( u\right) } \\ 
0 & 0 & y_{3}^{\left( u\right) }%
\end{array}%
\right) \frac{v}{\sqrt{2}},  \label{Mu} \\
M_{D} &=&\left( 
\begin{array}{ccc}
\varepsilon _{11}^{\left( d\right) }\lambda ^{4} & \varepsilon _{12}^{\left(
d\right) }\lambda ^{3} & 0 \\ 
\varepsilon _{21}^{\left( d\right) }\lambda ^{4} & \varepsilon _{22}^{\left(
d\right) }\lambda ^{3} & 0 \\ 
0 & 0 & y_{3}^{\left( d\right) }\lambda ^{3}%
\end{array}%
\right) \frac{v}{\sqrt{2}},  \label{Md}
\end{eqnarray}%
where $\varepsilon _{ij}^{\left( u,d\right) }$ 
are dimensionless parameters generated at one loop level 
whose corresponding expressions are given in Appendix \ref{ep}. In addition, $y_{3}^{\left( u\right) }$ and $y_{3}^{\left( d\right) }$ are $\mathcal{O}(1)$ dimensionless couplings generated at tree level from renormalizable and nonrenormalizable Yukawa terms, respectively. 
\begin{figure}[tbp]
\resizebox{20cm}{35cm}{\hspace{-2.5cm}\includegraphics[width=1\textwidth]{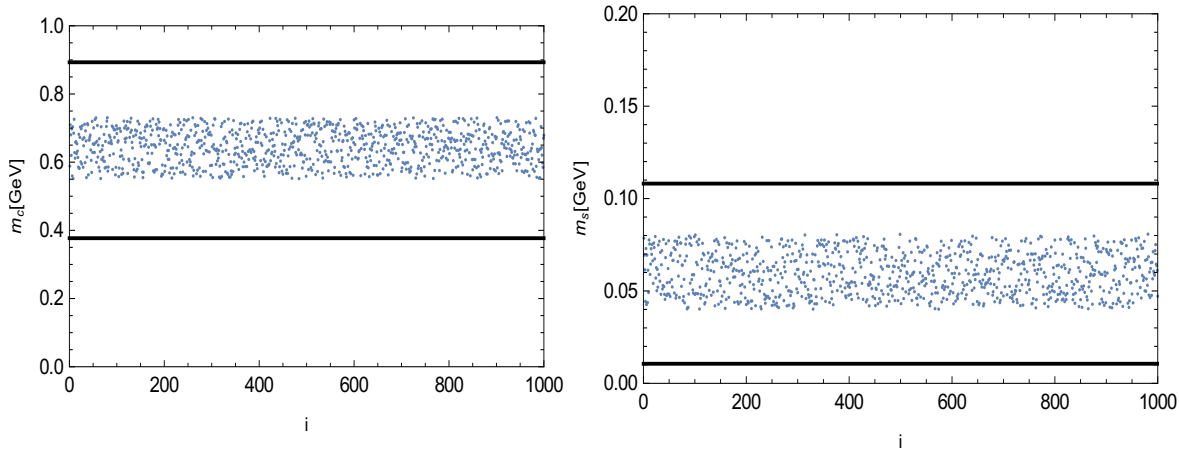}}
\vspace{-26cm}
\caption{Charm and strange quark masses randomly generated. The horizonal
lines are the minimum and maximum values of the SM quark masses inside the $3%
\protect\sigma $ experimentally allowed range.}
\label{mq}
\end{figure}
\begin{figure}[tbp]\vspace{-3cm}
\resizebox{20cm}{35cm}{\hspace{-2.5cm}\includegraphics[width=1\textwidth]{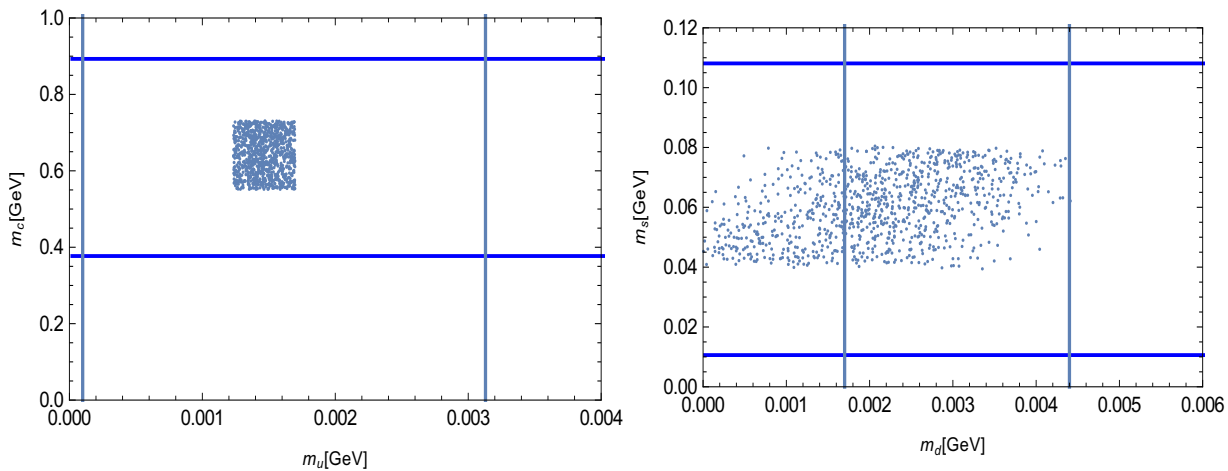}}
\vspace{-26cm}
\caption{Correlations between the first and second generation SM quark
masses. The horizonal and vertical lines are the minimum and maximum values
of the second and first generation quark masses, respectively, inside the $3%
\protect\sigma$ experimentally allowed range.}
\label{cq}
\end{figure}
%
In order to show that the quark textures given above can fit the
experimental data, and considering that the parameters $\varepsilon
_{ij}^{\left( u,d\right) }$ are generated at one loop level, we choose a
benchmark scenario where we set:
\begin{eqnarray}
\varepsilon _{11}^{\left( u\right) } &=&a_{11}^{\left( u\right) }\lambda
^{2},\hspace{1cm}\varepsilon _{22}^{\left( u\right) }=a_{22}^{\left(
u\right) }\lambda ^{2},\hspace{1cm}\varepsilon _{13}^{\left( u\right)
}=a_{13}^{\left( u\right) }\lambda ^{3},\hspace{1cm}\varepsilon
_{23}^{\left( u\right) }=a_{23}^{\left( u\right) }\lambda ^{2},  \notag \\
\varepsilon _{11}^{\left( d\right) } &=&a_{11}^{\left( d\right) }\lambda
^{3},\hspace{1cm}\varepsilon _{21}^{\left( d\right) }=a_{21}^{\left(
d\right) }\lambda ^{2},\hspace{1cm}\varepsilon _{12}^{\left( d\right)
}=a_{12}^{\left( d\right) }\lambda ^{3},\hspace{1cm}\varepsilon
_{22}^{\left( d\right) }=a_{22}^{\left( d\right) }\lambda ^{2}.  \label{ep2}
\end{eqnarray}%
where $a_{nn}^{\left(u\right)}$, $a_{n3}^{\left(u\right)}$, $a_{mn}^{\left(d\right)}$ ($m,n=1,2$) are $\mathcal{O}(1)$
parameters. Let us note that from the quark mass matrices given above, it
follows that the Cabbibo mixing arises from the down-type quark sector,
whereas the up-type quark sector contributes to the remaining mixing angles.
Besides that, the low energy quark flavor data indicates that the CP
violating phase in the quark sector is associated with the quark mixing
angle in the 1-3 plane, as follows from the Standard parametrization of the
quark mixing matrix. Consequently, in order to get quark mixing angles and a
CP violating phase consistent with the experimental data, we assume that all
dimensionless parameters given in Eqs. (\ref{Mu}) and (\ref{Md}) are real,
except $a_{13}^{\left( u\right) }$, which is taken to be complex. 
%
Since the observed pattern of charged fermion masses and quark mixing angles
is generated from the $S_{3}\otimes Z_{12}$ symmetry breaking, and in order
to have the right value of the Cabbibo mixing, we need 
$a_{21}^{\left(d\right) }\approx a_{22}^{\left( d\right) }$. In addition we set 
$y_{3}^{\left( u\right) }\approx 1$, as suggested by naturalness arguments.
Then the quark sector of our model contains nine effective free parameters,
i.e., $a_{11}^{\left( u\right) }$, $a_{22}^{\left( u\right) }$, 
$|a_{13}^{\left( u\right) }|$, $a_{23}^{\left( u\right) }$, $a_{11}^{\left(
d\right) }$, $a_{12}^{\left( d\right) }$, $a_{22}^{\left( d\right) }$, 
$y_{3}^{\left( d\right) }$ and the phase $\arg\left(a^{\left( u\right)}_{13}\right)$, which are fitted to
reproduce the ten physical observables of the quark sector, i.e., the six
quark masses, the three mixing angles and the CP violating phase. 
By varying these parameters, we find the quark masses, the three quark
mixing angles, and the CP violating phase $\delta $ reported in Table \ref%
{Quarkobs}, which correspond to the best-fit values: 
\begin{eqnarray}
a_{23}^{\left( u\right) } &\simeq &0.81,\hspace{1cm}|a_{13}^{\left( u\right)}|\simeq 0.3,\hspace{1cm}\arg\left(a^{\left( u\right) }_{13}\right) =-113^{\circ },\hspace{1cm}%
a_{22}^{\left( u\right) }\simeq 1.43,\hspace{1cm}a_{11}^{\left( u\right)
}\simeq 1.27,  \notag \\
a_{11}^{\left( d\right) } &\simeq &0.84,\hspace{1cm}a_{12}^{\left( d\right)
}\simeq 0.4,\hspace{1cm}a_{22}^{\left( d\right) }\simeq 0.57,\hspace{1cm}%
y_{3}^{\left( d\right) }\simeq 1.42.
\end{eqnarray}


\begin{table}[tbh]
\begin{center}
\begin{tabular}{c|l|l|}
\hline\hline
Observable & Model value & Experimental value \\ \hline
$m_{u}(MeV)$ & \quad $1.47$ & \quad $1.45_{-0.45}^{+0.56}$ \\ \hline
$m_{c}(MeV)$ & \quad $641$ & \quad $635\pm 86$ \\ \hline
$m_{t}(GeV)$ & \quad $172.2$ & \quad $172.1\pm 0.6\pm 0.9$ \\ \hline
$m_{d}(MeV)$ & \quad $3.00$ & \quad $2.9_{-0.4}^{+0.5}$ \\ \hline
$m_{s}(MeV)$ & \quad $59.2$ & \quad $57.7_{-15.7}^{+16.8}$ \\ \hline
$m_{b}(GeV)$ & \quad $2.82$ & \quad $2.82_{-0.04}^{+0.09}$ \\ \hline
$\sin \theta _{12}$ & \quad $0.2257$ & \quad $0.2254$ \\ \hline
$\sin \theta _{23}$ & \quad $0.0412$ & \quad $0.0413$ \\ \hline
$\sin \theta _{13}$ & \quad $0.00352$ & \quad $0.00350$ \\ \hline
$\delta $ & \quad $68^{\circ }$ & \quad $68^{\circ }$ \\ \hline\hline
\end{tabular}%
\end{center}
\caption{Model and experimental \cite{Olive:2016xmw,Bora:2012tx,Xing:2007fb} values of the quark masses and CKM
parameters.}
\label{Quarkobs}
\end{table}

It is worth mentioning, as follows from Eq. (\ref{epquarksector}) given in
Appendix \ref{ep}, that the functions $\varepsilon _{11}^{\left( u\right) }$%
, $\varepsilon _{22}^{\left( u\right) }$, $\varepsilon _{13}^{\left(
u\right) }$ and $\varepsilon _{23}^{\left( u\right) }$ depend on the
dimensionless parameters $x_{1}^{\left( u\right) }$, $x_{2}^{\left( u\right)
}$, $x_{3}^{\left( u\right) }$, $x_{4}^{\left( u\right) }$, $y_{1}^{\left(
u\right) }$, $y_{2}^{\left( u\right) }$, on the masses $m_{T_{1}}$, $%
m_{T_{2}}$, $m_{\func{Re}\phi _{2}^{0}}$, $m_{\func{Re}\eta }$, $m_{\func{Im}%
\phi _{2}^{0}}$ and $m_{\func{Im}\eta }$ as well as on the trilinear scalar
coupling $C_{\phi _{2}\phi _{1}\eta }$. \ Furthermore, the functions $%
\varepsilon _{11}^{\left( d\right) }$, $\varepsilon _{22}^{\left( d\right) }$%
, $\varepsilon _{12}^{\left( d\right) }$ and $\varepsilon _{21}^{\left(
d\right) }$ depend on the dimensionless parameters $x_{1}^{\left( d\right) }$%
, $x_{2}^{\left( d\right) }$, $x_{3}^{\left( d\right) }$, $x_{4}^{\left(
d\right) }$, $y_{1}^{\left( d\right) }$, $y_{2}^{\left( d\right) }$, on the
masses $m_{B_{1}}$, $m_{B_{2}}$, $m_{\func{Re}\phi _{2}^{0}}$, $m_{\func{Re}%
\eta }$, $m_{\func{Im}\phi _{2}^{0}}$ and $m_{\func{Im}\eta }$ as well as on
the trilinear scalar coupling $C_{\phi _{2}\phi _{1}\eta }$. Consequently,
there is a good amount of parametric freedom to reproduce the obtained
values for the $\varepsilon _{ij}^{\left( u,d\right) }$ ($i,j=1,2,3$)
functions that successfully reproduce the physical observables of the quark
sector. For instance, the best-fit values given above imply that $%
\varepsilon _{11}^{\left( u\right) }\simeq 6.4\times 10^{-2}$, which can be
obtained by setting $m_{T_{1}}=1$ TeV, $m_{\func{Re}\phi _{2}^{0}}=420$ GeV, 
$m_{\func{Re}\eta }=560$ GeV, $m_{\func{Im}\phi _{2}^{0}}=600$ GeV, $m_{%
\func{Im}\eta }=800$ GeV, $C_{\phi _{2}\phi _{1}\eta }=3$ TeV, $%
x_{3}^{\left( u\right) }=y_{1}^{\left( u\right) }=3$. Similar
considerations apply for the down-type quark and lepton sector.

In order to study the sensitivity of the obtained values for the SM quark
masses under small variations around the best-fit values (maximum variation
of $+0.2$, minimum of $-0.2$), we show in Fig.~\ref{mq} the predicted charm
and strange masses as functions of the iteration. We find that, for a slight
deviation from the best-fit values, the obtained charm and strange masses stay
inside the $3\sigma $ experimentally allowed range. We also have numerically
checked that the remaining SM quark masses are kept inside the $3\sigma $
experimentally allowed limits when we perform small variations around the
best-fit values, with the exception of the down and bottom quark masses.
The sensitivity of the down-type quark mass under small variations around the
best-fit values is due to the fact that all entries in the upper left block
of the SM down-type quark mass matrix are different from zero, thus implying
that there is an important region of parameter space where the determinant
of the down-type quark mass matrix takes low values. With respect to the
bottom quark mass, we find that most of the points are inside the $3\sigma $
experimentally allowed range. Those outside the $3\sigma $
experimentally allowed range correspond to values close to the lower and
upper experimental bounds of the bottom quark mass. Consequently, our model
is very predictive for the quark sector. Correlations between the first- and
second-generation SM quark masses are shown in Fig.~\ref{cq}. The horizontal
and vertical lines are the minimum and maximum values of the second- and
first- generation quark masses, respectively, inside the $3\sigma $
experimentally allowed range.


\section{Lepton masses and mixings}
%
\label{leptonmassesandmixings} 
From the lepton Yukawa terms (\ref{Lyl}) we derive 
the neutrino mass matrices
\begin{eqnarray}
M_{l} &=&\left( 
\begin{array}{ccc}
\varepsilon _{11}^{\left( l\right) }\lambda ^{6} & \varepsilon _{12}^{\left(
l\right) }\lambda ^{3} & 0 \\ 
\varepsilon _{21}^{\left( l\right) }\lambda ^{6} & \varepsilon _{22}^{\left(
l\right) }\lambda ^{3} & 0 \\ 
0 & 0 & y_{3}^{\left( l\right) }\lambda ^{3}%
\end{array}%
\right) \frac{v}{\sqrt{2}},  \label{Ml} \\
M_{\nu } &=&\left( 
\begin{array}{ccc}
W_{1}^{2} & W_{1}W_{2}\cos \varphi  & W_{1}W_{3}\cos \left( \varphi -\varrho
\right)  \\ 
W_{1}W_{2}\cos \varphi  & W_{2}^{2} & W_{2}W_{3}\cos \varrho  \\ 
W_{1}W_{3}\cos \left( \varphi -\varrho \right)  & W_{2}W_{3}\cos \varrho  & 
W_{3}^{2}%
\end{array}%
\right) ,  \label{Mnu}
\end{eqnarray}%
where $\varepsilon _{nm}^{\left( l\right) }$ ($n,m=1,2$) are dimensionless
parameters generated at one loop level whose corresponding expressions are
given in Appendix \ref{ep}. The parameters of the neutrino mass matrix of
Eq. (\ref{Mnu}) take the form
\begin{eqnarray}
\overrightarrow{W_{j}} &=&\left( \frac{A_{j1}\sqrt{m_{N_{1}}f_{1}^{\left(
\nu \right) }}}{64\pi ^{3}\Lambda },\frac{A_{j2}\sqrt{m_{N_{2}}f_{2}^{\left(
\nu \right) }}}{64\pi ^{3}\Lambda }\right) ,\hspace{0.5cm}j=1,2,3  \notag \\
\cos \varphi  &=&\frac{\overrightarrow{W_{1}}\cdot \overrightarrow{W_{2}}}{%
\left\vert \overrightarrow{W_{1}}\right\vert \left\vert \overrightarrow{W_{2}%
}\right\vert },\hspace{0.25cm}\cos \left( \varphi -\varrho \right) =\frac{%
\overrightarrow{W}_{1}\cdot \overrightarrow{W_{3}}}{\left\vert 
\overrightarrow{W}_{1}\right\vert \left\vert \overrightarrow{W_{3}}%
\right\vert },\hspace{0.25cm}W_{j}=\left\vert \overrightarrow{W_{j}}%
\right\vert ,  \notag \\
\cos \varrho  &=&\frac{\overrightarrow{W_{2}}\cdot \overrightarrow{W_{3}}}{%
\left\vert \overrightarrow{W_{2}}\right\vert \left\vert \overrightarrow{W_{3}%
}\right\vert },\hspace{0.5cm}A_{js}=y_{js}^{\left( \nu \right) }\frac{v}{%
\sqrt{2}}\hspace{0.5cm}s=1,2,  \notag \\
f_{s} &=&\frac{m_{R}^{2}}{m_{R}^{2}-m_{N_{s}}^{2}}\ln \left( \frac{m_{R}^{2}%
}{m_{N_{s}}^{2}}\right) -\frac{m_{I}^{2}}{m_{I}^{2}-m_{N_{s}}^{2}}\ln \left( 
\frac{m_{I}^{2}}{m_{N_{s}}^{2}}\right) ,  \notag \\
m_{R} &=&m_{\func{Re}\phi _{2}^{0}},\hspace{1cm}m_{I}=m_{\func{Im}\phi
_{2}^{0}}.
\end{eqnarray}

In order to show that the lepton textures given above can fit the
experimental data, and considering that the parameters $\varepsilon
_{ij}^{\left( l\right) }$ are generated at one loop level, we choose a
benchmark scenario where we set: 
\begin{equation}
\varepsilon _{11}^{\left( l\right) }=a_{11}^{\left( l\right) }\lambda ^{2},%
\hspace{1cm}\varepsilon _{21}^{\left( l\right) }=a_{21}^{\left( l\right)
}\lambda ^{2},\hspace{1cm}\varepsilon _{12}^{\left( l\right)
}=a_{12}^{\left( l\right) }\lambda ^{2},\hspace{1cm}\varepsilon
_{22}^{\left( l\right) }=a_{22}^{\left( l\right) }\lambda ^{2}.
\end{equation}%
where $a_{nm}^{\left( l\right) }$ ($n,m=1,2,3$) are $\mathcal{O}(1)$
parameters.

Then it follows that the charged lepton mass matrix takes the following
form: 
\begin{equation}
M_{l}=\frac{v}{\sqrt{2}}\left( 
\begin{array}{ccc}
a_{11}^{\left( l\right) }\lambda ^{8} & a_{12}^{\left( l\right) }\lambda ^{5}
& 0 \\ 
a_{21}^{\left( l\right) }\lambda ^{8} & a_{22}^{\left( l\right) }\lambda ^{5}
& 0 \\ 
0 & 0 & y_{3}^{\left( l\right) }\lambda ^{3}%
\end{array}%
\right) .
\end{equation}%
%
%
%
%
%
%
%
%
%
%
which implies that the following relation is fulfilled: 
\begin{equation}  \label{eq:M-l}
M_{l}M_{l}^{T}=\frac{v^{2}}{2}\left( 
\begin{array}{ccc}
\left( a_{12}^{\left( l\right) }\right) ^{2}\lambda ^{10}+\left(
a_{11}^{\left( l\right) }\right) ^{2}\lambda ^{16} & a_{12}^{\left( l\right)
}a_{22}^{\left( l\right) }\lambda ^{10}+a_{11}^{\left( l\right)
}a_{21}^{\left( l\right) }\lambda ^{16} & 0 \\ 
a_{12}^{\left( l\right) }a_{22}^{\left( l\right) }\lambda
^{10}+a_{11}^{\left( l\right) }a_{21}^{\left( l\right) }\lambda ^{16} & 
\left( a_{22}^{\left( l\right) }\right) ^{2}\lambda ^{10}+\left(
a_{21}^{\left( l\right) }\right) ^{2}\lambda ^{16} & 0 \\ 
0 & 0 & \left( y_{3}^{\left( l\right) }\right) ^{2}\lambda ^{6}%
\end{array}%
\right) \allowbreak .
\end{equation}

The mass matrix $M_{l}M_{l}^{T}$ can be diagonalized by a rotation matrix $%
R_{l}$ according to

\begin{equation}
R_{l}^{T}M_{l}M_{l}^{T}R_{l}=\left( 
\begin{array}{ccc}
m_{e}^{2} & 0 & 0 \\ 
0 & m_{\mu }^{2} & 0 \\ 
0 & 0 & m_{\tau }^{2}%
\end{array}%
\right) ,\hspace{1cm}\hspace{1cm}R_{l}=\left( 
\begin{array}{ccc}
\cos \theta _{l} & \sin \theta _{l} & 0 \\ 
-\sin \theta _{l} & \cos \theta _{l} & 0 \\ 
0 & 0 & 1%
\end{array}%
\right) \allowbreak ,\hspace{1cm}\hspace{1cm}\tan \theta _{l}\simeq \frac{%
a_{12}^{\left( l\right) }}{a_{22}^{\left( l\right) }},  \label{Rl}
\end{equation}

where the charged lepton masses are given by
\begin{eqnarray}
\label{mll-1}
m_{e} &\simeq &\lambda ^{8}\cdot 
\frac{\left|a_{12}^{\left( l\right)} a_{21}^{\left( l\right)} - a_{11}^{\left( l\right)} a_{22}^{\left( l\right)}\right|}
{\sqrt{\left( a_{12}^{\left( l\right) }\right) ^{2}+\left(a_{22}^{\left( l\right) }\right) ^{2}}}
\frac{v}{\sqrt{2}}, \\
\label{mll-2}
m_{\mu } &\simeq &\lambda ^{5}\cdot \sqrt{\left( a_{12}^{\left( l\right)
}\right) ^{2}+\left( a_{22}^{\left( l\right) }\right) ^{2}}
\frac{v}{\sqrt{2}},
%
\\
\label{mll-3}
m_{\tau } &=&\lambda ^{3}\cdot y_{3}^{\left( l\right) }\frac{v}{\sqrt{2}}.
\end{eqnarray}
Thus we correctly reproduce the charged lepton mass hierarchy from the
symmetry structure of the model. 

To simplify further the analysis, we set $\varphi =\varrho $, obtaining that
the light neutrino mass matrix is given by

\begin{equation}
M_{\nu }=\left( 
\begin{array}{ccc}
W_{1}^{2} & \kappa W_{1}W_{2} & W_{1}W_{3} \\ 
\kappa W_{1}W_{2} & W_{2}^{2} & \kappa W_{2}W_{3} \\ 
W_{1}W_{3} & \kappa W_{2}W_{3} & W_{3}^{2}%
\end{array}%
\right) ,\hspace{2cm}\kappa =\cos \varphi .  \label{Mnub}
\end{equation}

Assuming that the neutrino Yukawa couplings are real, we find that for the
normal (NH) and inverted (IH) mass hierarchies, the light neutrino mass
matrix is diagonalized by a rotation matrix $R_{\nu }$, according to
\begin{eqnarray}
 \label{NeutrinomassNH}
R_{\nu }^{T}M_{\nu }R_{\nu } &=&\left( 
\begin{array}{ccc}
0 & 0 & 0 \\ 
0 & m_{\nu _{2}} & 0 \\ 
0 & 0 & m_{\nu _{3}}%
\end{array}%
\right) \allowbreak ,\hspace{1cm}R_{\nu }=\left( 
\begin{array}{ccc}
-\frac{W_{3}}{\sqrt{W_{1}^{2}+W_{3}^{2}}} & \frac{W_{1}}{\sqrt{%
W_{1}^{2}+W_{3}^{2}}}\sin \theta _{\nu } & \frac{W_{1}}{\sqrt{%
W_{1}^{2}+W_{3}^{2}}}\cos \theta _{\nu } \\ 
0 & \cos \theta _{\nu } & -\sin \theta _{\nu } \\ 
\frac{W_{1}}{\sqrt{W_{1}^{2}+W_{3}^{2}}} & \frac{W_{3}}{\sqrt{%
W_{1}^{2}+W_{3}^{2}}}\sin \theta _{\nu } & \frac{W_{3}}{\sqrt{%
W_{1}^{2}+W_{3}^{2}}}\cos \theta _{\nu }%
\end{array}
\right) ,\hspace{5mm}\mbox{for NH}  \\
\label{eq:Tannu-NH}
\tan \theta _{\nu } &=&-\sqrt{\frac{W_{2}^{2}-m_{\nu _{2}}}{m_{\nu_{3}}-W_{2}^{2}}}\\
\label{eq:Mnu-NH}
m_{\nu _{1}}&=&0,\hspace{0.5cm}
m_{\nu _{2,3}}=\frac{W_{1}^{2}+W_{2}^{2}+W_{3}^{2}}{2}\mp \frac{\sqrt{\left(
W_{1}^{2}-W_{2}^{2}+W_{3}^{2}\right) ^{2}+4\kappa ^{2}W_{2}^{2}\left(
W_{1}^{2}+W_{3}^{2}\right) }}{2}. 
\end{eqnarray}
\begin{eqnarray}
\label{NeutrinomassIH}
R_{\nu }^{T}M_{\nu }R_{\nu } &=&\left( 
\begin{array}{ccc}
m_{\nu _{1}} & 0 & 0 \\ 
0 & m_{\nu _{2}} & 0 \\ 
0 & 0 & 0
\end{array}
\right) \allowbreak ,\hspace{1cm}R_{\nu }=\left( 
\begin{array}{ccc}
\frac{W_{1}}{\sqrt{W_{1}^{2}+W_{3}^{2}}}\sin \theta _{\nu } & \frac{W_{1}}{
\sqrt{W_{1}^{2}+W_{3}^{2}}}\cos \theta _{\nu } & -\frac{W_{3}}{\sqrt{W_{1}^{2}+W_{3}^{2}}} \\ 
\cos \theta _{\nu } & -\sin \theta _{\nu } & 0 \\ 
\frac{W_{3}}{\sqrt{W_{1}^{2}+W_{3}^{2}}}\sin \theta _{\nu } & \frac{W_{3}}{%
\sqrt{W_{1}^{2}+W_{3}^{2}}}\cos \theta _{\nu } & \frac{W_{1}}{\sqrt{%
W_{1}^{2}+W_{3}^{2}}}
\end{array}
\right) ,\hspace{5mm}\mbox{for IH}   \\
\label{eq:Tannu-IH}
\tan \theta _{\nu } &=&-\sqrt{\frac{W_{2}^{2}-m_{\nu _{1}}}{m_{\nu _{2}}-W_{2}^{2}}}\\
\label{eq:Mnu-IH}
m_{\nu _{1,2}}&=&\frac{W_{1}^{2}+W_{2}^{2}+W_{3}^{2}}{2}
\mp \frac{1}{2}\sqrt{\left(W_{1}^{2}-W_{2}^{2}+W_{3}^{2}\right) ^{2}+4\kappa ^{2}W_{2}^{2}
\left(W_{1}^{2}+W_{3}^{2}\right) },\hspace{5mm}m_{\nu _{3}}=0. 
\end{eqnarray}
%
With the rotation matrices in the charged lepton sector $R_{l}$, given by
Eq. (\ref{Rl}), and in the neutrino sector $R_{\nu }$, given by Eqs. (\ref%
{NeutrinomassNH}) and (\ref{NeutrinomassIH}) for NH and IH, respectively, we
find the PMNS mixing matrix: 
\begin{equation}
U=R_{l}^{T}R_{\nu }=\left\{ 
\begin{array}{l}
\left( 
\begin{array}{ccc}
-\frac{W_{3}}{\sqrt{W_{1}^{2}+W_{3}^{2}}}\cos \theta _{l} & \frac{W_{1}}{%
\sqrt{W_{1}^{2}+W_{3}^{2}}}\cos \theta _{l}\sin \theta _{\nu }-\cos \theta
_{\nu }\sin \theta _{l} & \sin \theta _{l}\sin \theta _{\nu }+\frac{W_{1}}{%
\sqrt{W_{1}^{2}+W_{3}^{2}}}\cos \theta _{l}\cos \theta _{\nu } \\ 
-\frac{W_{3}}{\sqrt{W_{1}^{2}+W_{3}^{2}}}\sin \theta _{l} & \cos \theta
_{l}\cos \theta _{\nu }+\frac{W_{1}}{\sqrt{W_{1}^{2}+W_{3}^{2}}}\sin \theta
_{\nu }\sin \theta _{l} & \frac{W_{1}}{\sqrt{W_{1}^{2}+W_{3}^{2}}}\sin
\theta _{l}\cos \theta _{\nu }-\cos \theta _{l}\sin \theta _{\nu } \\ 
\frac{W_{1}}{\sqrt{W_{1}^{2}+W_{3}^{2}}} & \frac{W_{3}}{\sqrt{%
W_{1}^{2}+W_{3}^{2}}}\sin \theta _{\nu } & \frac{W_{3}}{\sqrt{%
W_{1}^{2}+W_{3}^{2}}}\cos \theta _{\nu }%
\end{array}%
\right) \ \ \ \ \ \mbox{for NH}, \\ 
\left( 
\begin{array}{ccc}
\frac{W_{1}}{\sqrt{W_{1}^{2}+W_{3}^{2}}}\cos \theta _{l}\sin \theta _{\nu
}-\cos \theta _{\nu }\sin \theta _{l} & \sin \theta _{l}\sin \theta _{\nu }+%
\frac{W_{1}}{\sqrt{W_{1}^{2}+W_{3}^{2}}}\cos \theta _{l}\cos \theta _{\nu }
& -\frac{W_{3}}{\sqrt{W_{1}^{2}+W_{3}^{2}}}\cos \theta _{l} \\ 
\cos \theta _{l}\cos \theta _{\nu }+\frac{W_{1}}{\sqrt{W_{1}^{2}+W_{3}^{2}}}%
\sin \theta _{l}\sin \theta _{\nu } & \frac{W_{1}}{\sqrt{W_{1}^{2}+W_{3}^{2}}%
}\cos \theta _{\nu }\sin \theta _{l}-\cos \theta _{l}\sin \theta _{\nu } & -%
\frac{W_{3}}{\sqrt{W_{1}^{2}+W_{3}^{2}}}\sin \theta _{l} \\ 
\frac{W_{3}}{\sqrt{W_{1}^{2}+W_{3}^{2}}}\sin \theta _{\nu } & \frac{W_{3}}{%
\sqrt{W_{1}^{2}+W_{3}^{2}}}\cos \theta _{\nu } & \frac{W_{1}}{\sqrt{%
W_{1}^{2}+W_{3}^{2}}}%
\end{array}%
\right) \allowbreak \ \ \ \ \mbox{for IH}.%
\end{array}%
\right.
\end{equation}

From the standard parametrization of the leptonic mixing matrix, it follows
that the lepton mixing angles for NH and IH, respectively, are

\begin{eqnarray}
\sin ^{2}\theta _{12} &=&\frac{\left( \frac{W_{1}}{\sqrt{W_{1}^{2}+W_{3}^{2}}%
}\cos \theta _{l}\sin \theta _{\nu }-\cos \theta _{\nu }\sin \theta
_{l}\right) ^{2}}{1-\left( \sin \theta _{l}\sin \theta _{\nu }+\frac{W_{1}}{%
\sqrt{W_{1}^{2}+W_{3}^{2}}}\cos \theta _{l}\cos \theta _{\nu }\right) ^{2}},%
\hspace{1cm}\sin ^{2}\theta _{13}=\left( \sin \theta _{l}\sin \theta _{\nu }+%
\frac{W_{1}}{\sqrt{W_{1}^{2}+W_{3}^{2}}}\cos \theta _{l}\cos \theta _{\nu
}\right) ^{2},  \notag \\
\label{mixinganglesNH}
\sin ^{2}\theta _{23} &=&\frac{\left( \frac{W_{1}}{\sqrt{W_{1}^{2}+W_{3}^{2}}%
}\sin \theta _{l}\cos \theta _{\nu }-\cos \theta _{l}\sin \theta _{\nu
}\right) ^{2}}{1-\left( \sin \theta _{l}\sin \theta _{\nu }+\frac{W_{1}}{%
\sqrt{W_{1}^{2}+W_{3}^{2}}}\cos \theta _{l}\cos \theta _{\nu }\right) ^{2}}\
,\ \ \ \ \ \ \ \ \mbox{for NH} \\
%
\sin ^{2}\theta _{12} &=&\frac{\left( \sin \theta _{l}\sin \theta _{\nu }+%
\frac{W_{1}}{\sqrt{W_{1}^{2}+W_{3}^{2}}}\cos \theta _{l}\cos \theta _{\nu
}\right) ^{2}}{1-\frac{W_{3}^{2}\cos ^{2}\theta _{l}}{W_{1}^{2}+W_{3}^{2}}},%
\hspace{17mm}\sin ^{2}\theta _{13}=\frac{W_{3}^{2}\cos ^{2}\theta _{l}}{%
W_{1}^{2}+W_{3}^{2}},  \notag \\
 \label{mixingnaglesIH}
\sin ^{2}\theta _{23} &=&\frac{\frac{W_{3}^{2}}{W_{1}^{2}+W_{3}^{2}}\sin
^{2}\theta _{l}}{1-\frac{W_{3}^{2}\cos ^{2}\theta _{l}}{W_{1}^{2}+W_{3}^{2}}}
\ ,\hspace{56mm} \mbox{for IH} 
\end{eqnarray}
%
%
%
%
In the charged lepton sector the model has five free parameters $a_{11}^{(l)}, a_{12}^{(l)}, a_{21}^{(l)}, 
a_{22}^{(l)}, y_{3 }^{(l)}$. Fitting Eqs.~(\ref{mll-1})-(\ref{mll-3}) for $m_{e, \mu, \tau}$ to the corresponding experimental values \cite{Bora:2012tx}, we found $\mathcal{O}(1)$ solutions for these parameters. Therefore, we conclude that the model correctly predicts the charged lepton mass hierarchy according to 
Eqs.~(\ref{mll-1})-(\ref{mll-3}).  On the other hand, it does not predict the particular values of the charged lepton masses. In fact, within the hierarchical structure incorporated in Eqs.~(\ref{mll-1})-(\ref{mll-3}), 
the five free $\mathcal{O}(1)$ parameters $a_{11}^{(l)}, a_{12}^{(l)}, a_{21}^{(l)},
a_{22}^{(l)}, y_{3 }^{(l)}$ allow to reproduce the experimental
values of $m_{e,\mu,\tau}$ with arbitrary precision limited only by the experimental
errors.
%

\begin{figure}[tbp]
\vspace{-3cm}
\resizebox{20cm}{35cm}{\hspace{-2.5cm}\includegraphics[width=1\textwidth]{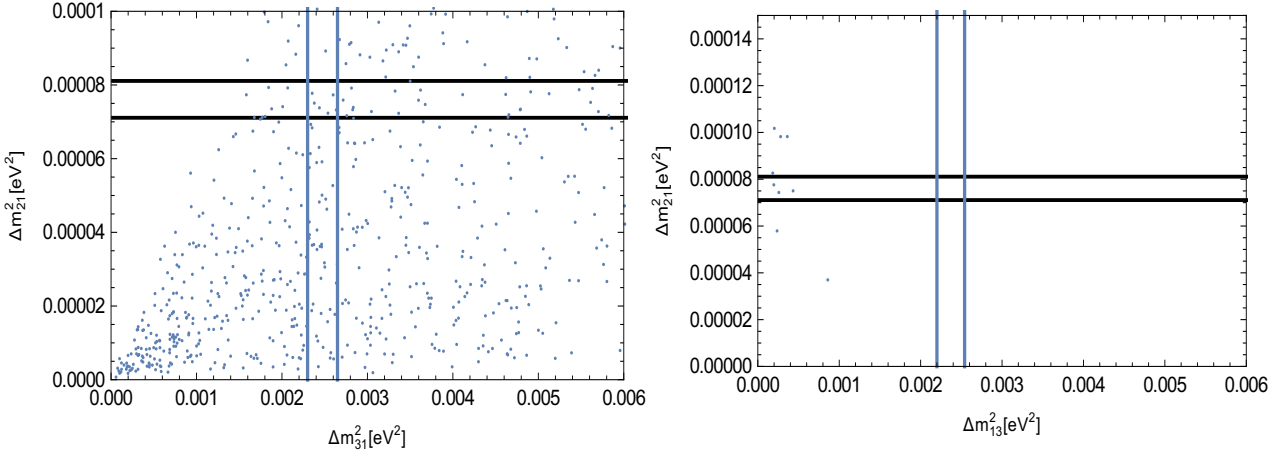}}
\vspace{-26cm}
\caption{Correlations between $\Delta m_{21}^{2}$ and $\Delta m_{31}^{2}$
for the cases of normal (left plot) and inverted (right plot) neutrino mass
hierarchy. The horizonal and vertical lines are the minimum and maximum
values of the neutrino mass squared splittings inside the $3\protect\sigma$
experimentally allowed range.}
\label{Correlationneutrinos}
\end{figure}

The charged lepton sector \textquotedblleft contributes\textquotedblright\
to the neutrino sector with a single free parameter $\theta _{l}$ defined in
Eq.~(\ref{Rl}). It remains unrestricted by the charged lepton masses. 
Thus, in the neutrino sector we have five free parameters 
$W_{1,2,3},\kappa ,\theta _{l}$ 
and five observables: two neutrino mass squared splittings $\Delta
m_{21}^{2}$, $\Delta m_{31}^{2}$ (we define $\Delta
m_{ij}^{2}=m_{i}^{2}-m_{j}^{2}$) and three mixing angles 
$\sin^{2}\theta _{12}$, $\sin ^{2}\theta _{13}$, $\sin ^{2}\theta _{23}$ for
both NH and IH. We solve Eqs.~(\ref{eq:Mnu-NH}), (\ref{eq:Mnu-IH}) and (\ref{mixinganglesNH}),
(\ref{mixingnaglesIH}) with respect to the model parameters for the central values of the corresponding observables shown in Table~\ref{Observables0}.
%
The unique solutions for NH and IH are 
\begin{eqnarray}
\mbox{for NH:} &&\kappa \simeq 0.7,\hspace{15mm}W_{1}\simeq 0.09\,\mbox{eV}^{%
\frac{1}{2}},\ \ W_{2}\simeq 0.16\mbox{eV}^{\frac{1}{2}},\ \ W_{3}\simeq
0.15\,\mbox{eV}^{\frac{1}{2}},\ \ \theta _{l}\simeq 18.95^{\circ }
\label{ParameterfitNH} \\
\mbox{for}\ \ \mbox{IH:} &&\kappa \simeq 6.11\times 10^{-3},\ \ W_{1}\simeq
0.14\,\mbox{eV}^{\frac{1}{2}},\ \ W_{2}\simeq 0.22\,\mbox{eV}^{\frac{1}{2}%
},\ \ W_{3}\simeq 0.17\,\mbox{eV}^{\frac{1}{2}},\ \ \theta _{l}\simeq
-78.30^{\circ }.  \label{ParameterfitIH}
\end{eqnarray}
Fig.~\ref{Correlationneutrinos} shows the correlation between $\Delta m_{21}^{2}$ and 
$\Delta m_{31}^{2}$ for the cases of normal and inverted
neutrino mass hierarchies, respectively. We find that a slight variation
from the best-fit values yields, for several points of the parameter space,
an important deviation in the values of the neutrino mass squared splittings
and leptonic mixing parameters, especially for the case of inverted neutrino
mass hierarchy.

%
%
%
\begin{table}[tbh]
\begin{center}
\begin{tabular}{c|l|c|c|c}
\hline\hline
Observable & Experimental value&\hspace{3mm}{\bf (i)}&{\bf (ii)'}&{\bf (ii)}\\ \hline
$m_{e}(MeV)$ & \quad $0.487$ &{\rm no prediction}&{\rm input}&0.43\\ 
\hline
$m_{\mu }(MeV)$ &\quad $102.8\pm 0.0003$& {\rm no prediction}& 113.07 &115.1 \\ 
%
\hline
$\Delta m_{21}^{2}$($10^{-5}$eV$^{2}$) (NH) &  
\quad $7.60_{-0.18}^{+0.19}$&7.62\hspace*{5mm}&7.72 &7.62\\ 
\hline
$\Delta m_{31}^{2}$($10^{-3}$eV$^{2}$) (NH) & \quad $2.48_{-0.07}^{+0.05}$ 
&2.52\hspace*{5mm} & 2.70 &2.52\\ 
\hline
$\sin ^{2}\theta _{12}$ (NH) & \quad $0.323\pm 0.016$&0.324\hspace*{3.5mm}&0.258 &0.303\\ 
\hline
$\sin ^{2}\theta _{23}$ (NH) & \quad $0.567_{-0.128}^{+0.032}$&0.542\hspace*{3.5mm}&0.526 &0.527\\ 
\hline
$\sin ^{2}\theta _{13}$ (NH) & \quad $0.0234\pm 0.0020$&0.0234\hspace*{2mm}& 0.0224&0.0204\\ 
\hline
$\Delta m_{21}^{2}$($10^{-5}$eV$^{2}$) (IH) & \quad $7.60_{-0.18}^{+0.19}$ 
&-&-&-\\ 
\hline
$\Delta m_{13}^{2}$($10^{-3}$eV$^{2}$) (IH) & \quad $2.38_{-0.06}^{+0.05}$ &-&-&-\\ 
\hline
$\sin ^{2}\theta _{12}$ (IH) & \quad $0.323\pm 0.016$ &-&-&-\\ 
\hline
$\sin ^{2}\theta _{23}$ (IH) & \quad $0.573_{-0.043}^{+0.025}$&-& -&-\\ 
\hline
$\sin ^{2}\theta _{13}$ (IH) & \quad $0.0240\pm 0.0019$&-& -&-\\ 
\hline
\end{tabular}
\end{center}
\par
\caption{The symmetry inspired benchmark scenarios {\bf (i)} and {\bf (ii)} versus  
the experimental values of the charged lepton masses \cite{Bora:2012tx},
neutrino oscillation observables for the
normal (NH) and inverted (IH) mass hierarchies \cite{Forero:2014bxa}. 
Scenario {\bf (i)}:  the four model parameters (\ref{NHmutau1}) fit the five neutrino oscillation observables for the NH.
Scenario {\bf (ii)'}:  from the experimental input value $m_{e}=0.487$MeV we determine a model parameter  $a_{12}^{(l)}=0.52$ and then calculate $m_{\mu}=113.07$MeV; the neutrino oscillation observables for the NH are fitted with three parameters (\ref{eq:ii-1}). Scenario {\bf (ii)'}: the same as {\bf (ii)}, but fitting jointly the four model parameters (\ref{resultsNHp}) to the seven observables.
}
\label{Observables0}
\end{table}
Nevertheless the model in its present version is not predictive in the lepton sector except for the charged lepton mass hierarchy Eqs.~(\ref{mll-1})-(\ref{mll-3}).  This is because the number of free parameters equals the number of observables. 

However, considering the model parameters for NH in Eq.~ (\ref{ParameterfitNH}) we note that their numerical values $W_{2}\simeq W_{3}$, $\kappa \approx 1$ may point to an approximate underlying $\mu -\tau $ symmetry. In fact for $W_{2}=W_{3}$, $\kappa =1$ the light active neutrino mass matrix (\ref{Mnub})  coincides in structure with 
the $\mu -\tau $ symmetric Fukuyama-Nishiura texture \cite{Fukuyama:1997ky}:
%
%
\begin{equation}
M_{\nu }=\left( 
\begin{array}{ccc}
A & B & B \\ 
B & C & D \\ 
B & D & C%
\end{array}%
\right) .  \label{FKN}
\end{equation}%
%
With this observation let us study a $\mu -\tau$-symmetry inspired benchmark scenario in our model. 
Incorporation of the $\mu -\tau$-symmetry in our model implies $W_{2}=W_{3}$
and $\kappa =1$ and, therefore, $C=D$ in (\ref{FKN}) leading to two
vanishing eigenvalues, which is in clear contradiction with the 
neutrino oscillation data. After all this symmetry cannot be exact, since it is explicitly broken in 
the charged lepton sector (for more details on the phenomenological aspects of the $\mu -\tau $ interchange symmetry see, for instance, Refs.~ \cite{Mohapatra:2004mf}). 
However, the $\mu-\tau$-symmetry can be incorporated in our model as a ``minimally'' broken symmetry. 
Its maximal breaking corresponds to the violation of both $W_{2}=W_{3}$
and $\kappa =1$ conditions. In this sense the minimal breaking is introduced by the violation of only one of these conditions. As can easily be checked the case $W_{2}\neq W_{3}$ with $\kappa =1$ is an inappropriate one since it has two zero eigenvalues of the neutrino mass matrix Eq.~(\ref{Mnub}). Thus, we study the benchmark scenario\\
{\bf (i)} with   $W_{2}= W_{3}$ and $\kappa$ being a free $\mu-\tau$-symmetry breaking parameter. In this case there is only one zero neutrino mass matrix eigenvalue. Now we have four free model parameters 
%
$W_{1}$, $W_{2}$, $\kappa $ and 
$\theta _{l}$ versus the five physical
observables in the neutrino sector, i.e., the two neutrino mass squared
splittings and the three leptonic mixing parameters. Thus, the model is able to predict one of these observables using the experimental values of the other four. Instead of doing this we conventionally fit the model parameters to the experimental values of  all the five observables within the corresponding experimental errors. 
%
The best fit results are shown in Table~\ref{Observables0} for 
\begin{eqnarray}
\kappa  &\simeq &0.70,\hspace{1cm}W_{1}\simeq 0.09\mbox{eV}^{\frac{1}{2}},%
\hspace{1cm}W_{2}\simeq 0.16\mbox{eV}^{\frac{1}{2}},\hspace{1cm}\theta
_{l}\simeq 19.81^{\circ }.  \label{NHmutau1}
\end{eqnarray}
%
corresponding to the normal neutrino hierarchy. Here the parameter $\kappa$ is not too far from its
$\mu-\tau$-symmetric value  $\kappa=1$. In the case of the inverted hierarchy it is two-three 
orders of magnitude away from from this value and  the $\mu-\tau$-symmetry does not show up 
as an approximate symmetry of the neutrino sector.  A more detailed study of this 
observation and its possible implementation in the model Lagrangian will be addressed elsewhere.   
%

Let us examine even more restrictive benchmark scenario
assuming additionally to $W_{3}=W_{2}$ also the relation for the model parameters 
$a_{11}^{\left( l\right) }=a_{22}^{\left( l\right)}=a_{21}^{\left( l\right) }=1$
in the sector of charged leptons $e, \mu$.  So that there is only one free parameter $a_{12}^{\left( l\right)}$ 
to fit $m_{e}$ and $m_{\mu}$ using Eqs.~(\ref{mll-1})-(\ref{mll-2}).  This scenario is motivated by the fact we already 
discussed above (see the paragraph after Eqs.~(\ref{mixingnaglesIH})) that our model correctly reproduces the charged lepton mass hierarchy with 
all the parameters $a_{11}^{(l)}, a_{12}^{(l)}, a_{21}^{(l)}, a_{22}^{(l)}$ of the order $\mathcal{O}(1)$. 
Setting all these parameters to be $a_{ij}^{(l)} = 1$ leads to a rather rough estimate of $m_{e}$ and $m_{\mu}$.
Moreover, in this case $\tan\theta_{l} =1$, which is incompatible with the neutrino oscillation data as seen from (\ref{NHmutau1}). Therefore, we release $a_{12}^{(l)}$ assuming it to be the only free parameter in 
the $e-\mu$-sector of charged leptons. Thus we consider a benchmark scenario\\
{\bf (ii)} with $a_{11}^{\left( l\right) }=a_{22}^{\left( l\right)}=a_{21}^{\left( l\right) }=1$, $W_{2}=W_{3}$.
There are only four free parameters are $W_{1}, W_{2}, \kappa$ and  $a_{12}^{(l)}$ to reproduce seven observable: $m_{e}, m_{\mu}$ and five neutrino oscillation parameters. Note that the charged lepton 
$e-\mu$-sector depends only on one of these parameters, namely,  $a_{12}^{(l)}$.  It is tempting to try to make a prediction for $m_{\mu}$ starting from the best measured $m_{e}$. This also restricts the number of the free parameters in the neutrino sector down to $W_{1}, W_{2}, \kappa$. In this way we solve Eq.~(\ref{mll-1}) with respect to $a_{12}^{(l)}$ for the central experimental value of $m_{e}$  and predict $m_{\mu}$ using (\ref{mll-2}). The result is shown in Table~\ref{Observables0} in the column  (ii)' for $a_{12}^{(l)}=0.52$ obtained from $m_{e}$. 
In the neutrino sector, having three parameters, the model is able to predict 
any two of the five observables using  as an input the experimental values of the other three.  However, as in the scenario (i) we fit all the neutrino oscillation observables, 
$\Delta m_{21}^{2}$, $\Delta m_{31}^{2}$, $\sin^{2}\theta _{12}$, $\sin ^{2}\theta _{13}$, 
$\sin ^{2}\theta _{23}$ with the model parameters $W_{1}, W_{2}, \kappa$. The best fit values shown in the column {\bf (ii)'} of Table~\ref{Observables0} correspond to 
\begin{eqnarray} 
\label{eq:ii-1}
&& \kappa=0.71,\ \ \ W_{1}=0.05\mbox{eV}^{\frac{1}{2}},\ \ \ W_{2}=-0.17\mbox{eV}^{\frac{1}{2}}. 
\end{eqnarray}
Now, instead of fixing $a_{12}^{(l)}$ from $m_{e}$, we fit
all the model parameters $W_{1}, W_{2}, \kappa, a_{12}^{(l)}$ to seven observables $m_{e}, m_{\mu}$ and $\Delta m_{21}^{2}$, $\Delta m_{31}^{2}$, $\sin^{2}\theta _{12}$, $\sin ^{2}\theta _{13}$, 
$\sin ^{2}\theta _{23}$. The best-fit values are shown in the column (ii) of Table~\ref{Observables0} and correspond to the following values of the model parameters:
\begin{eqnarray}
\kappa  &\simeq &0.71,\hspace{1cm}W_{1}\simeq 0.06\mbox{eV}^{\frac{1}{2}},%
\hspace{1cm}W_{2}\simeq - 0.17\mbox{eV}^{\frac{1}{2}},\hspace{1cm}%
a_{12}^{\left( l\right) }\simeq 0.56.  \label{resultsNHp}
\end{eqnarray}%
As seen from Table~\ref{NHexp} in the scenario (ii) the neutrino oscillation parameters 
$\sin ^{2}\theta _{12}$, $\sin ^{2}\theta _{23}$, $\Delta m_{21}^{2}$, $\Delta m_{31}^{2}$
are inside the $1\sigma $ experimentally allowed range. The
reactor mixing parameter $\sin ^{2}\theta _{13}$ is 
inside the $2\sigma$
range. 


Now we evaluate in our model the effective Majorana neutrino mass parameter
of neutrinoless double beta ($0\nu \beta \beta $) decay
\begin{equation}
m_{\beta \beta }=\left\vert \sum_{j}U_{ek}^{2}m_{\nu _{k}}\right\vert ,
\label{mee}
\end{equation}%
where $U_{ej}^{2}$ and $m_{\nu _{k}}$ are the PMNS mixing matrix elements
and the Majorana neutrino masses, respectively.

\begin{table}[tbh]
\begin{tabular}{|c|c|c|c|c|c|}
\hline
Parameter & $\Delta m_{21}^{2}$($10^{-5}$eV$^2$) & $\Delta m_{31}^{2}$($%
10^{-3}$eV$^2$) & $\left( \sin ^{2}\theta _{12}\right) _{\exp }$ & $\left(
\sin ^{2}\theta _{23}\right) _{\exp }$ & $\left( \sin ^{2}\theta
_{13}\right) _{\exp }$ \\ \hline
Best fit & $7.60$ & $2.48$ & $0.323$ & $0.567$ & $0.0234$ \\ \hline
$1\sigma $ range & $7.42-7.79$ & $2.41-2.53$ & $0.307-0.339$ & $0.439-0.599$
& $0.0214-0.0254$ \\ \hline
$2\sigma $ range & $7.26-7.99$ & $2.35-2.59$ & $0.292-0.357$ & $0.413-0.623$
& $0.0195-0.0274$ \\ \hline
$3\sigma $ range & $7.11-8.11$ & $2.30-2.65$ & $0.278-0.375$ & $0.392-0.643$
& $0.0183-0.0297$ \\ \hline
\end{tabular}%
\caption{Range for experimental values of neutrino mass squared splittings
and leptonic mixing parameters, taken from Ref.~\protect\cite{Forero:2014bxa}, 
for the case of normal hierarchy.}
\label{NHexp}
\end{table}

As we already mentioned the model parameters in Eq.~(\ref{Mnu}) may have arbitrary complex phases. Therefore there may occur cancellation between the terms in Eq.~(\ref{mee}), which we are not able to control in the model.  Thus we can predict only upper bounds for $m_{\beta\beta}$. For the parameters (\ref{ParameterfitNH}), (\ref{ParameterfitIH}) we find
\begin{equation}
m_{\beta \beta }\leq\left\{ 
\begin{array}{l}\ \, 
4\ \mbox{meV}\ \ \ \ \  \ \  \mbox{for \ \ \ \ NH} \\ 
49\ \mbox{meV}\ \ \ \ \ \ \ \mbox{for \ \ \ \ IH} \\ 
\end{array}%
\right.  \label{eff-mass-pred}
\end{equation}%
%
%
%
%
%
%
%
%
%
%
%
%
%
%
%
%
%
%
%
%
%
%
%
%
%
%
%
%
%
%
%
%
%
%
%
%
%
%
%
%
%
%
%
%
%
%
%
%
%
%
Our obtained value $m_{\beta \beta }\approx 4\ \mbox{meV}$ for the effective
Majorana neutrino mass parameter is beyond the reach of the present and
forthcoming $0\nu \beta \beta $ decay experiments. The current best upper
bound on the effective neutrino mass is $m_{\beta \beta }\leq 160$ meV,
which corresponds to $T_{1/2}^{0\nu \beta \beta }(^{136}\mathrm{Xe})\geq
1.1\times 10^{26}$ years at 90\% C.L, as indicated by the KamLAND-Zen
experiment \cite{KamLAND-Zen:2016pfg}. This bound will be improved within a
not too far future. The GERDA \textquotedblleft phase-II\textquotedblright
experiment \cite{Abt:2004yk,Ackermann:2012xja} 
is expected to reach 
\mbox{$T^{0\nu\beta\beta}_{1/2}(^{76}{\rm Ge}) \geq
2\times 10^{26}$ years}, which corresponds to $m_{\beta \beta }\leq 100$ meV. A
bolometric CUORE experiment, using ${}^{130}Te$ \cite{Alessandria:2011rc},
is currently under construction and has an estimated sensitivity of about $%
T_{1/2}^{0\nu \beta \beta }(^{130}\mathrm{Te})\sim 10^{26}$ years, which
corresponds to \mbox{$m_{\beta\beta}\leq 50$ meV.} Furthermore, there are
proposals for ton-scale next-to-next generation $0\nu \beta \beta $
experiments with $^{136}$Xe \cite{KamLANDZen:2012aa,Albert:2014fya} and $%
^{76}$Ge \cite{Abt:2004yk,Guiseppe:2011me}, claiming sensitivities over $%
T_{1/2}^{0\nu \beta \beta }\sim 10^{27}$ years, which corresponds to $m_{\beta
\beta }\sim 12-30$ meV. For a recent review, see for example Ref. \cite%
{Bilenky:2014uka}. Consequently, as follows from Eq. (\ref{eff-mass-pred}),
our model predicts $T_{1/2}^{0\nu \beta \beta }$ at the level of
sensitivities of the next generation or next-to-next generation $0\nu \beta
\beta $ experiments.\newline

\section{Dark Matter candidate}
\label{sec:DM}
%
It is worth mentioning that there is a viable dark matter (DM) candidate 
in our model. If we consider a specific scenario where only the extra scalar
doublet $\phi_{2}$ acquires mass at low energy, while the rest of extra
scalars and exotic fermions live at high energies, an approach to the well-know inert Higgs doublet model could be done. This model, originally
proposed in \cite{Deshpande:1977rw} and extensively studied in a number of
recent works \cite{Honorez:2010re, Diaz:2015pyv}, introduces an additional
doublet, namely denoted in the literature as $H_{2}$, odd under an
additional $Z_{2}$ symmetry. The lightest inert $Z_{2}$-odd particle turns out to
be stable and hence a suitable DM candidate. Our additional scalar doublet $%
\phi_{2}$, analogous to $H_{2}$, is odd under $Z_{2}$, which guarantees that
it does not have direct couplings with SM fermion pairs and then its
stability. The inert Higgs dark matter, in particular, could annihilate into 
$WW^*$, $ZZ^*$, $hh^* $ and $t\overline{t}^*$. In the region restricted by
the relic density constraint, i.e: $\Omega_{DM} h^2=0.1181 \pm 0.0012$ the
annihilation on the DM into $W^+$ $W^-$ and $ZZ$ is very effective, much
more than annihilation into $hh^*$ and $t\overline{t}^*$. The DM constraints
can only be satisfied for restricted values of $m_{\phi_{2}}$. In \cite%
{Diaz:2015pyv}, three viable regions of DM are pointed out: a small regime
with $3\leq m_{H_{2}} \leq 50$ GeV, an intermediate regime with $60\leq
m_{H_{2}} \leq 100$ GeV and a large regime $3 m_{H_{2}} \geq 550$ GeV.
Therefore, if we expect $\phi_{2}$ to be a DM candidate, its mass must fall
in some of the three allowed regions. The limits on $m_{\phi_{2}}$ can be
translated into some restrictions on the other model parameters, for
example, $\epsilon_{ij}^{f}$ ($f=u,d,\nu$), whose precise values are fixed
by the requirement of having a realistic spectrum of SM fermion masses and
mixing angles. The resulting constraints on the $\epsilon_{ij}^{l}$
parameters will yield bounds on the exotic fermion masses, thus setting
limits on the total production cross sections of the non-SM particles at the
LHC. Derivation of these constraints requires a dedicated study beyond the
scope of the present paper and is left for future studies. 

\section{Conclusions}

\label{conclusions} We have constructed an extension of the inert 2HDM,
based on the extended $S_{3}\otimes Z_{2}\otimes Z_{12}$ symmetry that
successfully describes the current pattern of SM fermion masses and mixings.
In our model the $Z_{2}$ is preserved, whereas the $S_{3}$ and $Z_{12}$
symmetries are broken, giving rise to the observed pattern of charged
fermion masses and mixing angles. The preserved $Z_{2}$ symmetry allows the
implementation of a one loop level radiative seesaw mechanism, which
generates the masses for the first- and second-generation charged fermions,
as well as a non-trivial quark mixing. 

In our modified inert 2HDM, the SM fermion masses and mixing pattern arise
from a combination of tree and one-loop-level effects. At tree level only the
third-generation charged fermions acquire masses and there is no quark
mixing, while the first- and second-generation charged fermion masses and the
quark mixing arise from one loop level radiative seesaw-type mechanisms,
triggered by virtual $Z_{2}$-charged scalar fields and electrically charged
exotic fermions running inside the loops. Light active neutrino masses are
generated from a one-loop-level radiative seesaw mechanism.

As follows from the expressions for the loop functions given in Eq.~(\ref%
{loopfunction}), the entries of the SM charged fermion and light active
neutrino mass matrices generated at one-loop-level are monotonically
decreasing functions of the masses of the other particles in the loops shown
in Fig.~\ref{loopdiagrams}. 
The condition that the mass matrices be compatible with the observed masses
and mixing of the SM fermions sets constraints on the masses of the $Z_{2}$%
-odd scalars and the non-SM fermions. Derivation of these constraints
requires a dedicated study because of the complexity of the model parameter
space. %
%


The preserved $Z_{2}$ symmetry of our model allows not only implementation
of the radiative seesaw-type mechanism, but also provides natural dark
matter candidates, stable due to this symmetry. 
They are the right-handed Majorana neutrinos $N_{1R}$, $N_{2R}$ and/or the
lightest of the $Z_{2}$ odd scalars $\func{Re}\eta $, $\func{Im}\eta $, $%
\func{Re}\phi _{2}^{0}$, $\func{Im}\phi _{2}^{0}$. Their masses are
constrained by the dark matter relic density. The constraints on the masses
of the $Z_{2}$ odd scalars states, will yield bounds on the total production
cross sections of these particles at the LHC. The implications of our model
in collider physics and dark matter requires careful studies that we left
outside the scope of this paper and defer for a future publication. 
\newline
\textbf{Acknowledgments}. 
\newline
This work was partially supported by Fondecyt (Chile), Grants No.~1170803,
No.~1150792, No.~1140390, No.~3150472 and by CONICYT (Chile) Ring ACT1406
and CONICYT PIA/Basal FB0821.

\section*{Appendices}

\appendix


\section{The product rules of the $S_{3}$ discrete group}

\label{S3} The $S_{3}$ group has three irreducible representations: $\mathbf{%
1}$, $\mathbf{1}^{\prime }$ and $\mathbf{2}$. Denoting the basis vectors for
two $S_{3}$ doublets as $\left( x_{1},x_{2}\right) ^{T}$\ and $\left(
y_{1},y_{2}\right) ^{T}$ and $y^{\prime }$ a non-trivial $S_{3}$ singlet,
the multiplication rules for the $S_{3}$ discrete group take the form \cite%
{Ishimori:2010au}: 
\begin{equation}
\left( 
\begin{array}{c}
x_{1} \\ 
x_{2}%
\end{array}%
\right) _{\mathbf{2}}\otimes \left( 
\begin{array}{c}
y_{1} \\ 
y_{2}%
\end{array}%
\right) _{\mathbf{2}}=\left( x_{1}y_{1}+x_{2}y_{2}\right) _{\mathbf{1}%
}+\left( x_{1}y_{2}-x_{2}y_{1}\right) _{\mathbf{1}^{\prime }}+\left( 
\begin{array}{c}
x_{2}y_{2}-x_{1}y_{1} \\ 
x_{1}y_{2}+x_{2}y_{1}%
\end{array}%
\right) _{\mathbf{2}},  \label{6}
\end{equation}%
\begin{equation}
\left( 
\begin{array}{c}
x_{1} \\ 
x_{2}%
\end{array}%
\right) _{\mathbf{2}}\otimes \left( y^{\prime }\right) _{\mathbf{1}^{\prime
}}=\left( 
\begin{array}{c}
-x_{2}y^{\prime } \\ 
x_{1}y^{\prime }%
\end{array}%
\right) _{\mathbf{2}},\hspace{1cm}\hspace{1cm}\left( x^{\prime }\right) _{%
\mathbf{1}^{\prime }}\otimes \left( y^{\prime }\right) _{\mathbf{1}^{\prime
}}=\left( x^{\prime }y^{\prime }\right) _{\mathbf{1}}.  \label{7}
\end{equation}

\section{Analytical expressions for the dimensionless parameters of the SM
fermion mass matrices generated at one loop level}

\label{ep} The dimensionless parameters $\varepsilon _{ij}^{\left(
u,d\right) }$ ($i,j=1,2,3$) generated at one loop level that appear in the
SM quark mass matrices are given by
\begin{eqnarray}
\varepsilon _{11}^{\left( u\right) } &=&\frac{1}{16\pi ^{2}}\frac{C_{\phi
_{2}\phi _{1}\eta }x_{3}^{\left( u\right) }y_{1}^{\left( u\right) }}{%
m_{T_{1}}}\left[ C_{0}\left( \frac{m_{\func{Re}\phi _{2}^{0}}}{m_{T_{1}}},%
\frac{m_{\func{Re}\eta }}{m_{T_{1}}}\right) -C_{0}\left( \frac{m_{\func{Im}%
\phi _{2}^{0}}}{m_{T_{1}}},\frac{m_{\func{Im}\eta }}{m_{T_{1}}}\right) %
\right] ,  \notag \\
\varepsilon _{22}^{\left( u\right) } &=&\frac{1}{16\pi ^{2}}\frac{C_{\phi
_{2}\phi _{1}\eta }x_{4}^{\left( u\right) }y_{2}^{\left( u\right) }}{%
m_{T_{2}}}\left[ C_{0}\left( \frac{m_{\func{Re}\phi _{2}^{0}}}{m_{T_{2}}},%
\frac{m_{\func{Re}\eta }}{m_{T_{2}}}\right) -C_{0}\left( \frac{m_{\func{Im}%
\phi _{2}^{0}}}{m_{T_{2}}},\frac{m_{\func{Im}\eta }}{m_{T_{2}}}\right) %
\right] ,  \notag \\
\varepsilon _{13}^{\left( u\right) } &=&\frac{1}{16\pi ^{2}}\frac{C_{\phi
_{2}\phi _{1}\eta }x_{1}^{\left( u\right) }y_{1}^{\left( u\right) }}{%
m_{T_{1}}}\left[ C_{0}\left( \frac{m_{\func{Re}\phi _{2}^{0}}}{m_{T_{1}}},%
\frac{m_{\func{Re}\eta }}{m_{T_{1}}}\right) -C_{0}\left( \frac{m_{\func{Im}%
\phi _{2}^{0}}}{m_{T_{1}}},\frac{m_{\func{Im}\eta }}{m_{T_{1}}}\right) %
\right] ,  \notag \\
\varepsilon _{23}^{\left( u\right) } &=&\frac{1}{16\pi ^{2}}\frac{C_{\phi
_{2}\phi _{1}\eta }x_{2}^{\left( u\right) }y_{2}^{\left( u\right) }}{%
m_{T_{2}}}\left[ C_{0}\left( \frac{m_{\func{Re}\phi _{2}^{0}}}{m_{T_{2}}},%
\frac{m_{\func{Re}\eta }}{m_{T_{2}}}\right) -C_{0}\left( \frac{m_{\func{Im}%
\phi _{2}^{0}}}{m_{T_{2}}},\frac{m_{\func{Im}\eta }}{m_{T_{2}}}\right) %
\right] ,  \notag \\
\varepsilon _{11}^{\left( d\right) } &=&\frac{1}{16\pi ^{2}}\frac{C_{\phi
_{2}\phi _{1}\eta }x_{1}^{\left( d\right) }y_{1}^{\left( d\right) }}{%
m_{B_{1}}}\left[ C_{0}\left( \frac{m_{\func{Re}\phi _{2}^{0}}}{m_{B_{1}}},%
\frac{m_{\func{Re}\eta }}{m_{B_{1}}}\right) -C_{0}\left( \frac{m_{\func{Im}%
\phi _{2}^{0}}}{m_{B_{1}}},\frac{m_{\func{Im}\eta }}{m_{B_{1}}}\right) %
\right] ,  \notag \\
\varepsilon _{22}^{\left( d\right) } &=&\frac{1}{16\pi ^{2}}\frac{C_{\phi
_{2}\phi _{1}\eta }x_{4}^{\left( d\right) }y_{2}^{\left( d\right) }}{%
m_{B_{2}}}\left[ C_{0}\left( \frac{m_{\func{Re}\phi _{2}^{0}}}{m_{B_{2}}},%
\frac{m_{\func{Re}\eta }}{m_{B_{2}}}\right) -C_{0}\left( \frac{m_{\func{Im}%
\phi _{2}^{0}}}{m_{B_{2}}},\frac{m_{\func{Im}\eta }}{m_{B_{2}}}\right) %
\right] ,  \notag \\
\varepsilon _{21}^{\left( d\right) } &=&\frac{1}{16\pi ^{2}}\frac{C_{\phi
_{2}\phi _{1}\eta }x_{3}^{\left( d\right) }y_{2}^{\left( d\right) }}{%
m_{B_{2}}}\left[ C_{0}\left( \frac{m_{\func{Re}\phi _{2}^{0}}}{m_{B_{2}}},%
\frac{m_{\func{Re}\eta }}{m_{B_{2}}}\right) -C_{0}\left( \frac{m_{\func{Im}%
\phi _{2}^{0}}}{m_{B_{2}}},\frac{m_{\func{Im}\eta }}{m_{B_{2}}}\right) %
\right] ,  \notag \\
\varepsilon _{12}^{\left( d\right) } &=&\frac{1}{16\pi ^{2}}\frac{C_{\phi
_{2}\phi _{1}\eta }x_{2}^{\left( d\right) }y_{1}^{\left( d\right) }}{%
m_{B_{1}}}\left[ C_{0}\left( \frac{m_{\func{Re}\phi _{2}^{0}}}{m_{B_{1}}},%
\frac{m_{\func{Re}\eta }}{m_{B_{1}}}\right) -C_{0}\left( \frac{m_{\func{Im}%
\phi _{2}^{0}}}{m_{B_{1}}},\frac{m_{\func{Im}\eta }}{m_{B_{1}}}\right) %
\right] ,  \label{epquarksector}
\end{eqnarray}%
where the following function has been introduced: 
\begin{equation}
C_{0}\left( \widehat{m}_{1},\widehat{m}_{2}\right) =\frac{1}{\left( 1-%
\widehat{m}_{1}^{2}\right) \left( 1-\widehat{m}_{2}^{2}\right) \left( 
\widehat{m}_{1}^{2}-\widehat{m}_{2}^{2}\right) }\left\{ \widehat{m}_{1}^{2}%
\widehat{m}_{2}^{2}\ln \left( \frac{\widehat{m}_{1}^{2}}{\widehat{m}_{2}^{2}}%
\right) -\widehat{m}_{1}^{2}\ln \widehat{m}_{1}^{2}+\widehat{m}_{2}^{2}\ln 
\widehat{m}_{2}^{2}\right\}   \label{loopfunction}
\end{equation}%
The dimensionless parameters $\varepsilon _{nm}^{\left( l\right) }$ ($n,m=1,2
$) generated at one loop level that appear in the SM charged lepton mass
matrices take the form
\begin{eqnarray}
\varepsilon _{11}^{\left( l\right) } &=&\frac{C_{\phi _{2}\phi _{1}\eta }}{%
16\pi ^{2}}\left\{ \frac{x_{1}^{\left( l\right) }y_{1}^{\left( l\right) }}{%
m_{E_{1}}}\left[ C_{0}\left( \frac{m_{\func{Re}\phi _{2}^{0}}}{m_{E_{1}}},%
\frac{m_{\func{Re}\eta }}{m_{E_{1}}}\right) -C_{0}\left( \frac{m_{\func{Im}%
\phi _{2}^{0}}}{m_{E_{1}}},\frac{m_{\func{Im}\eta }}{m_{E_{1}}}\right) %
\right] \right.   \notag \\
&&+\left. \frac{x_{3}^{\left( l\right) }y_{3}^{\left( l\right) }}{m_{E_{2}}}%
\left[ C_{0}\left( \frac{m_{\func{Re}\phi _{2}^{0}}}{m_{E_{2}}},\frac{m_{%
\func{Re}\eta }}{m_{E_{2}}}\right) -C_{0}\left( \frac{m_{\func{Im}\phi
_{2}^{0}}}{m_{E_{2}}},\frac{m_{\func{Im}\eta }}{m_{E_{2}}}\right) \right]
\right\}   \notag \\
\varepsilon _{22}^{\left( l\right) } &=&\frac{C_{\phi _{2}\phi _{1}\eta }}{%
16\pi ^{2}}\left\{ \frac{x_{2}^{\left( l\right) }y_{2}^{\left( l\right) }}{%
m_{E_{1}}}\left[ C_{0}\left( \frac{m_{\func{Re}\phi _{2}^{0}}}{m_{E_{1}}},%
\frac{m_{\func{Re}\eta }}{m_{E_{1}}}\right) -C_{0}\left( \frac{m_{\func{Im}%
\phi _{2}^{0}}}{m_{E_{1}}},\frac{m_{\func{Im}\eta }}{m_{E_{1}}}\right) %
\right] \right.   \notag \\
&&+\left. \frac{x_{4}^{\left( l\right) }y_{4}^{\left( l\right) }}{m_{E_{2}}}%
\left[ C_{0}\left( \frac{m_{\func{Re}\phi _{2}^{0}}}{m_{E_{2}}},\frac{m_{%
\func{Re}\eta }}{m_{E_{2}}}\right) -C_{0}\left( \frac{m_{\func{Im}\phi
_{2}^{0}}}{m_{E_{2}}},\frac{m_{\func{Im}\eta }}{m_{E_{2}}}\right) \right]
\right\}   \notag \\
\varepsilon _{21}^{\left( l\right) } &=&\frac{C_{\phi _{2}\phi _{1}\eta }}{%
16\pi ^{2}}\left\{ \frac{x_{2}^{\left( l\right) }y_{1}^{\left( l\right) }}{%
m_{E_{1}}}\left[ C_{0}\left( \frac{m_{\func{Re}\phi _{2}^{0}}}{m_{E_{1}}},%
\frac{m_{\func{Re}\eta }}{m_{E_{1}}}\right) -C_{0}\left( \frac{m_{\func{Im}%
\phi _{2}^{0}}}{m_{E_{1}}},\frac{m_{\func{Im}\eta }}{m_{E_{1}}}\right) %
\right] \right.   \notag \\
&&+\left. \frac{x_{4}^{\left( l\right) }y_{3}^{\left( l\right) }}{m_{E_{2}}}%
\left[ C_{0}\left( \frac{m_{\func{Re}\phi _{2}^{0}}}{m_{E_{2}}},\frac{m_{%
\func{Re}\eta }}{m_{E_{2}}}\right) -C_{0}\left( \frac{m_{\func{Im}\phi
_{2}^{0}}}{m_{E_{2}}},\frac{m_{\func{Im}\eta }}{m_{E_{2}}}\right) \right]
\right\}   \notag \\
\varepsilon _{12}^{\left( l\right) } &=&\frac{C_{\phi _{2}\phi _{1}\eta }}{%
16\pi ^{2}}\left\{ \frac{x_{1}^{\left( l\right) }y_{2}^{\left( l\right) }}{%
m_{E_{1}}}\left[ C_{0}\left( \frac{m_{\func{Re}\phi _{2}^{0}}}{m_{E_{1}}},%
\frac{m_{\func{Re}\eta }}{m_{E_{1}}}\right) -C_{0}\left( \frac{m_{\func{Im}%
\phi _{2}^{0}}}{m_{E_{1}}},\frac{m_{\func{Im}\eta }}{m_{E_{1}}}\right) %
\right] \right.   \notag \\
&&+\left. \frac{x_{3}^{\left( l\right) }y_{4}^{\left( l\right) }}{m_{E_{2}}}%
\left[ C_{0}\left( \frac{m_{\func{Re}\phi _{2}^{0}}}{m_{E_{2}}},\frac{m_{%
\func{Re}\eta }}{m_{E_{2}}}\right) -C_{0}\left( \frac{m_{\func{Im}\phi
_{2}^{0}}}{m_{E_{2}}},\frac{m_{\func{Im}\eta }}{m_{E_{2}}}\right) \right]
\right\} 
\end{eqnarray}

\end{document}